\documentclass[10pt]{elsarticle0}
\usepackage{epsfig}
\usepackage{slashed}
\usepackage{amsmath}
\usepackage{amssymb}

\newcommand{\beqn}{\begin{eqnarray}}
\newcommand{\eeqn}{\end{eqnarray}}

\usepackage{color}

\newcommand{\cor}[1]{{#1}}

\begin{document}

\title{Wigner transformation, momentum space topology, and anomalous transport}

\author[LESTUDIUM,LMPT,ITEP,DFU,MEPHI]{M.A.~Zubkov
\footnote{
e-mail: zubkov@itep.ru, on leave of absence from Moscow Institute of Physics and Technology, 9, Institutskii per., Dolgoprudny, Moscow Region, 141700, Russia}
 }

\address[LESTUDIUM]{ LE STUDIUM, Loire Valley Institute for Advanced Studies,
Tours and Orleans, 45000 Orleans France}

\address[LMPT]{Laboratoire de Math´ematiques et de Physique
Th´eorique, Universit´e de Tours, 37200 Tours, France  }


\address[ITEP]{ITEP, B.Cheremushkinskaya 25, Moscow, 117259, Russia
}


\address[DFU]{Far Eastern Federal University,  School of Biomedicine, 690950 Vladivostok, Russia}

\address[MEPHI]{National Research Nuclear University MEPhI (Moscow Engineering
Physics Institute), Kashirskoe highway 31, 115409 Moscow, Russia}

\begin{abstract}
Using derivative expansion applied to the Wigner transform of the two - point Green function we analyse the
anomalous quantum Hall effect (AQHE), and the chiral magnetic effect (CME). The corresponding currents are proportional to the momentum space topological invariants. We reproduce the conventional expression for the Hall conductivity in $2+1$ D. In $3+1$ D our analysis allows to explain systematically the AQHE in topological insulators and Weyl semimetals. At the same time using this method it may be proved, that the equilibrium CME is absent in the wide class of solids, as well as in the properly regularized relativistic quantum field theory.
\end{abstract}



\maketitle

\section{Introduction}

Momentum space topology is becoming the important tool for the study of the ground states of condensed matter systems (for the review see
\cite{HasanKane2010,Xiao-LiangQi2011,Volovik2011,Volovik2007,Volovik2010}). In particular, the momentum space topological invariants protect gapless fermions on the boundaries of topological insulators \cite{Gurarie2011,EssinGurarie2011}. Topological invariants in momentum space protect also the bulk gapless fermions in Dirac and Weyl semi - metals \cite{Volovik2003,VolovikSemimetal}. The large variety of topological defects and textures exist in the fermionic superfluids, and the gapless fermions associated with these objects are described by momentum space topology \cite{Volovik2016}. Momentum space topology was also discussed in the context of relativistic quantum field theory (QFT) \cite{NielsenNinomiya1981,So1985,IshikawaMatsuyama1986,Kaplan1992,Golterman1993,Volovik2003,Horava2005,Creutz2008,Kaplan2011}. In \cite{VZ2012} the topological invariants in momentum space have been considered for the lattice regularization of QFT with Wilson fermions. Appearance of the massless fermions at the intermediate values of bare mass parameter was related to the jump of the introduced momentum space topological invariant. This invariant may actually be used for the description of a certain class of topological insulators\footnote{This pattern is complimentary to that of \cite{EssinGurarie2011}, where in the similar way the number of boundary gapless fermions is related to the jump of the topological invariant across the boundary in coordinate space.}. In \cite{Z2012} the model with overlap fermions has been considered on the same grounds. In particular, the possible physical meaning of the zeros of the Green function has been discussed. The appearance of zeros of the Green function, in turn, has been discussed in the context of condensed matter physics (see, for example, \cite{zeros}).

The momentum space topological invariants are expressed in terms of the Green functions. Therefore, they are applicable both to the non - interacting and to the interacting systems \cite{Volovik2003}. Suppose, that we start from the model without interactions. When the interactions are turned on, the value of the topological invariant is not changed until the phase transition is encountered. This means, that the properties of the system described by the given topological invariant are robust to the introduction of interactions. The more simple non - interacting model may be investigated in order to describe such properties of the complicated interacting system. In the present paper we apply momentum space topology to the description of the anomalous quantum Hall effect (AQHE) in topological insulators and Weyl semimetals. In \cite{HAL} we considered the chiral magnetic effect (CME) (mainly, in the framework of relativistic quantum field theory) using the approach based on momentum space topology. Here we briefly repeat our proof of the absence of the equilibrium bulk CME with the emphasis in the application to the solid state systems.

Actually, momentum space topology with topological invariants expressed through the Green functions represents the alternative to the less powerful but more popular technique of Berry curvature proposed to describe the QHE in \cite{TTKN} and developed later in a number of publications (see, for example, \cite{Hall3DTI} and references therein). The technique of Berry curvature may be applied to the noninteracting condensed matter systems with  Green function of the form ${\cal G}^{-1} = i \omega - \hat{H}$, where $\omega$ is the imaginary frequency while $\hat{H}$ is the Hamiltonian.  Unfortunately, this formalism does not allow to deal in a similar way with the interacting systems with more complicated dependence of the Green function on the imaginary frequency. Besides, in spite of all its advantages, the Berry curvature formalism does not allow to describe the response of the system to the external magnetic (rather than electric) field. Finally, the bulk - boundary correspondence remains out of this formalism. Those three points are improved in the technique that utilizes the topological invariants composed directly of the Green functions. It was proposed first by G.E.Volovik (see \cite{Volovik2003,Gurarie2011} and references therein). Following \cite{HAL} in the present paper we develop this technique and give the relation between the electromagnetic response of electric current and the topological invariants in momentum space of both $2+1$D and $3+1$D  systems. In the present paper we emphasise, that those topological invariants are constructed of the Wigner transform of the two - point Green functions, which allows to apply the proposed methodology to the direct description of bulk - boundary correspondence. As for the Berry curvature formalism, we demonstrate that it follows from our approach as a particular case.

The family of the non - dissipative transport effects related to chiral anomaly has been widely discussed recently both in the context of the high energy physics and in the context of condensed matter theory \cite{CME,ZrTe5,SonYamamoto2012,Landsteiner:2012kd,semimetal_effects7,Gorbar:2015wya,Miransky:2015ava,Valgushev:2015pjn,Buividovich:2015ara,Buividovich:2015ara,Buividovich:2014dha,Buividovich:2013hza}. In particular, the possible appearance of such effects in the recently discovered Dirac and Weyl semimetals has been considered \cite{semimetal_effects6,semimetal_effects10,semimetal_effects11,semimetal_effects12,semimetal_effects13,Zyuzin:2012tv,tewary}.
Besides, the possibility to observe those effects in relativistic heavy - ion collisions was proposed \cite{Kharzeev:2015znc,Kharzeev:2009mf}. The chiral magnetic effect (CME) is the generation of electric current in the presence of external magnetic field and chiral chemical potential \cite{Nielsen:1983rb,Kharzeev:2009pj}. The quantum Hall effect in the $2+1$ D and the $3+1$ D systems is the appearance of electric current in the direction orthogonal to the direction of the applied electric field. The anomalous quantum Hall effect (AQHE) is such an effect that takes place due to the internal properties of the system rather than due to the external magnetic field. Recently its possible appearance in Weyl semimetals has been widely discussed \cite{Zyuzin:2012tv,tewary}. The AQHE was also discussed in the $3+1$D topological insulators \cite{Hall3DTI}.  In the present paper we consider AQHE within the lattice regularized quantum field theory and within the tight - binding models of the solid state physics. The expression of the Hall current (through the topological invariant in momentum space $\tilde{\cal N}_3$) in the quasi two dimensional condensed matter systems  is well - known \cite{Volovik2003,Volovik1988}. We reproduce this result basing on the technique of the derivative expansion applied to the Wigner transform of the two - point Green function. Next, we apply the same technique to the AQHE in the $3+1$ D lattice models. The obtained expression relates the AQHE current to the new topological invariant. The technique of its calculation is developed. Our methodology allows to predict the appearance of the AQHE both in the Weyl semi - metals and in certain topological insulators.

We represent here the consideration of the equilibrium CME, which was based on the same technique. It appears, that the resulting current is also proportional to the topological invariant in momentum space. Unlike the case of the naive continuum fermions for the considered lattice models with regular fermionic Green function the value of the mentioned topological invariant does not depend on the chiral chemical potential. We consider the case, when the lattice Dirac fermions are massive. However, the limit, when the physical Dirac mass tends to zero, does not change our conclusion. This indicates, that the equilibrium bulk CME current is absent for the lattice regularized quantum field theory and in the certain class of solids.

Notice, that our conclusions on AQHE and CME are in accordance with the recent consideration of the particular lattice model of Weyl semimetals \cite{nogo}. The conclusion on the absence of the equilibrium bulk CME is in line with the recent numerical calculations made using the particular lattice models \cite{Valgushev:2015pjn,Buividovich:2015ara,Buividovich:2015ara,Buividovich:2014dha,Buividovich:2013hza}. It also does not contradict to the consideration of the effective continuum field theoretic description of Weyl/Dirac semimetals \cite{Gorbar:2015wya} and the consideration of Dirac semimetals in the framework of the semi - classical approach \cite{semiclassnoCME}. Besides, this conclusion is in accordance with the Bloch theorem \cite{nogo2}.

We incorporate the slowly varying external gauge field directly to the momentum space representation of the lattice model\cite{HAL}. In this representation it appears as a pseudo - differential operator ${\bf A}(i\partial_{\bf p})$, where  ${\bf A}({\bf r})$ determines the functional dependence of the gauge field on coordinates, while $\partial_{\bf p}$ is the derivative over momentum. Such an introduction of the gauge field to the lattice model deforms it at small distances (of the order of the lattice spacing). However, the gauge invariance remains intact, and at large distances the theory is indistinguishable from the lattice model with the external gauge field introduced in a conventional way. Therefore, the given way to introduce the electromagnetic field satisfies all requirements to be imposed on the lattice regularization of the QFT. We also consider this appropriate to apply this method to the lattice models of solid state physics if the long range phenomena are to be investigated. We demonstrate explicitly, that in the particular case of the noninteracting condensed matter system our formalism allows to reproduce the  expression for the Hall conductivity as an integral over the Berry curvature (the Green function of such system has the form ${\cal G}^{-1} = i \omega - \hat{H}$, where $\omega$ is the imaginary frequency while $\hat{H}$ is the Hamiltonian).

The paper is organized as follows. In Sections \ref{SectGaugeLat}, \ref{SectWignLat}, \ref{SectLinLat} we describe the formalism proposed to be used \cite{HAL}.  In Sect. \ref{SectGaugeLat} we describe the way to introduce the slowly varying external gauge field to the momentum space formulation of lattice model. In Sect. \ref{SectWignLat} we consider the Wigner transformation of the two - point Green function in momentum space. The linear response of electric current to external field strength is considered in Sect. \ref{SectLinLat}. In Sect. \ref{SectHall} we reproduce the expression for the QHE current in the $2+1$ D systems using the developed methodology. In Sect. \ref{SectBerry} we demonstrate, how the  formalism of Berry curvature follows from our expressions. In Sect. \ref{SectHall3d} we consider the AQHE in the $3+1$D systems. In Sect. \ref{bulkboundary} we discuss bulk - boundary correspondence for the topological insulators with AQHE. The considerations of Sect. \ref{SectHall3d} are illustrated by two particular models of insulators in Sect. \ref{SectIns}, and by two particular models of Weyl semimetal in Sect. \ref{SectWeylSem}. In Sect. \ref{SectCME} we analyse using this methodology the CME, and demonstrate, that it is absent in the considered systems. In Sect. \ref{SectConcl} we end with the conclusions.

\section{External gauge field as a pseudo - differential operator in momentum space}
\label{SectGaugeLat}

We consider the $d+1 = D$ dimensional lattice model of non - interacting fermions. In this way the tight - binding models of solid state physics as well as the lattice regularization of the continuum quantum field theory (QFT) may be described. In momentum space the fermionic Green function ${\cal G}({\bf p})$ depends on the $D$ vector ${\bf p} = (p_1,...,p_D)$ of Euclidean momentum. Momentum space is supposed to have the form of the product ${\cal M} = S^1 \otimes \Omega$ or ${\cal M} = R^1 \otimes \Omega$, where $\Omega$ is the compact $d$ - dimensional Brillouin zone. Here $R^1$ is the line of the imaginary frequencies $p^D$. In the lattice regularization of QFT the values of $p^D$ correspond to $S^1$. In condensed matter physics the representation of the theory with momentum space ${\cal M} = S^1 \otimes \Omega$ appears if the evolution in time is discretized. In particular, momentum space of such form appears in the application of the numerical lattice methods to the calculation of the functional integrals in the solid state physics. In the conventional lattice models defined on the hypercubic lattice $\Omega$ has the form of the $d$ - dimensional torus\footnote{Notice, that the lattice momentum ${\bf p}$ does not appear as the eigenvalue of the operator $-i\partial_{\bf r}$.}. In the solid state physics the form of the Brillouin zone is typically more complicated.

In the absence of the external electromagnetic field  the partition function of the theory defined on the infinite lattice may be written as
\begin{equation}
Z = \int D\bar{\Psi}D\Psi \, {\rm exp}\Big( - \int_{\cal M} \frac{d^D {\bf p}}{|{\cal M}|} \bar{\Psi}({\bf p}){\cal G}^{-1}({\bf p})\Psi({\bf p}) \Big)\label{Z1}
\end{equation}
where $|{\cal M}|$ is the volume of momentum space $\cal M$, while $\bar{\Psi}$ and $\Psi$ are the Grassmann - valued fields defined in momentum space $\cal M$. For example, the model with $3+1$ D Wilson fermions that describes qualitatively a certain class of topological insulators corresponds to $\cal G$ that has the form
 \begin{equation}
 {\cal G}({\bf p}) = \Big(\sum_{k}\gamma^{k} g_{k}({\bf p}) - i m({\bf p})\Big)^{-1}\label{G10}
 \end{equation}
 where $\gamma^k$ are Euclidean Dirac matrices while $g_k({\bf p})$ and $m({\bf p})$ are the real - valued functions ($k = 1,2,3,4$) given by
\begin{equation}
g_k({\bf p}) = {\rm sin}\,
p_k, \quad m({\bf p}) = m^{(0)} +
\sum_{a=1,2,3,4} (1 - {\rm cos}\, p_a)\label{gWilson}
\end{equation}
The fields in coordinate space are related to the fields in momentum space as follows
\begin{equation}
\Psi({\bf r}) = \int_{\cal M} \frac{d^D {\bf p}}{|{\cal M}|} e^{i {\bf p}{\bf r}} \Psi({\bf p})\label{Psip}
\end{equation}
At the discrete values of $\bf r$ corresponding to the points of the lattice this expression gives the values of the fermionic field at these points, i.e. the dynamical variables of the original lattice model. However, Eq. (\ref{Psip}) allows to define formally the values of fields at any other values of $\bf r$. The partition function may be rewritten in the form
\begin{equation}
Z = \int D\bar{\Psi}D\Psi \, {\rm exp}\Big( - \sum_{{\bf r}_n} \bar{\Psi}({\bf r}_n)\Big[{\cal G}^{-1}(-i\partial_{\bf r})\Psi({\bf r})\Big]_{{\bf r}={\bf r}_n} \Big)\label{Z2}
\end{equation}
Here the sum in the exponent is over the discrete coordinates ${\bf r}_n$. However, the operator $-i\partial_{\bf r}$ acts on the function $\Psi({\bf r})$ defined using Eq. (\ref{Psip}). In order to derive Eq. (\ref{Z2}) we use identity
\begin{equation}
\sum_{\bf r}e^{i{\bf p}{\bf r}} = |{\cal M}|\delta({\bf p})
\end{equation}
In the particular case of Wilson fermions\footnote{Wilson fermions  are widely use in the lattice discretization of QFT and also describe qualitatively a certain class of topological insulators.} we may rewrite the partition function in the conventional way as
\begin{equation}
Z = \int D\bar{\Psi}D\Psi \, {\rm exp}\Big( - \sum_{{\bf r}_n,{\bf r}_m} \bar{\Psi}({\bf r}_m)(-i{\cal D}_{{\bf r}_n,{\bf r}_m}) \Psi({\bf r}_n) \Big)\label{Z28}
\end{equation}
with
\begin{equation}
{\cal D}_{{\bf x},{\bf y}}  =  - \frac{1}{2}\sum_i [(1 +
\gamma^i)\delta_{{\bf x}+{\bf e}_i, {\bf y}}  +  (1 -
\gamma^i)\delta_{{\bf x}-{\bf e}_i, {\bf y}} ] +  (m^{(0)} + 4)
\delta_{{\bf x}{\bf y}}
\end{equation}
Here ${\bf e}_i$ is the unity vector in the $i$ - th direction.
Gauge transformation of the lattice field takes the form
\begin{equation}
\Psi({\bf r}_n)\rightarrow e^{i \alpha({\bf r}_n)} \Psi({\bf r}_n)\label{gt}
\end{equation}
In case of Wilson fermions the $U(1)$ gauge field is typically introduced as the following modification of operator $D$:
\begin{equation}
{\cal D}_{{\bf x},{\bf y}}  =  - \frac{1}{2}\sum_i [(1 +
\gamma^i)\delta_{{\bf x}+{\bf e}_i, {\bf y}}e^{i A_{{\bf x}+{\bf e}_i, {\bf y}}}  +  (1 -
\gamma^i)\delta_{{\bf x}-{\bf e}_i, {\bf y}} e^{i A_{{\bf x}-{\bf e}_i, {\bf y}}}] +  (m^{(0)} + 4)
\delta_{{\bf x}{\bf y}}
\end{equation}
Here $A_{{\bf x}, {\bf y}} = - A_{{\bf y}, {\bf x}}$ is the gauge field attached to  the links of the lattice. In the same way the gauge field is typically incorporated into the models of solid state physics.

Below we propose the alternative way of the introduction of gauge field into the lattice model.
First of all, Eq. (\ref{gt}) may be understood as the gauge transformation of the field $\Psi$ defined for any values of $\bf r$: we simply extend the definition of the function $\alpha({\bf r})$ to the function, which is defined at any values of $\bf r$ and take the original values at the discrete lattice points. This prompts the following way to introduce the external gauge field to our lattice model. Suppose, that we need to put on the lattice the gauge field, which in continuum theory has the form of the function ${\bf A}({\bf r})$ of the continuum coordinate $\bf r$.  We consider the partition function of the form
\begin{eqnarray}
Z &=& \int D\bar{\Psi}D\Psi \, {\rm exp}\Big(-\frac{1}{2}\sum_{{\bf r}_n}\Big[ \bar{\Psi}({\bf r}_n)\Big[{\cal G}^{-1}(-i\partial_{\bf r}  \nonumber\\&& - {\bf A}({\bf r}))\Psi({\bf r})\Big]_{{\bf r}={\bf r}_n} + (h.c.)\Big]\Big)\label{Z3}
\end{eqnarray}
Here by $(h.c.)$ we denote the Hermitian conjugation, which is defined as follows. First of all, it relates the components of Grassmann variable $\Psi$ with the corresponding components of $\bar{\Psi}$. Besides, it inverses the ordering of operators and the variables $\bar{\Psi},\Psi$, and substitutes each operator by its Hermitian conjugated. For example, a conjugation of $\bar{\Psi} \hat{B} (i \partial_{r^{i_1}})...(i \partial_{r^{i_n}}) \Psi$ for a certain operator (in internal space) $\hat{B}$  is given by $\Big[(-i \partial_{r^{i_1}})...(-i \partial_{r^{i_n}}) \bar{\Psi} \Big]\hat{B}^+ \Psi$. As well as in continuum theory operators $\hat{p}_i - A_i({\bf r})$ and $\hat{p}_j - A_j({\bf r})$ do not commute for $i \ne j$. Therefore, we should point out the way of their ordering inside ${\cal G}^{-1}(-i\partial_{\bf r} - {\bf A}({\bf r}))$. We choose the following way for definiteness:  each product $p_{i_1} ... p_{i_n}$ in the expansion of ${\cal G}^{-1}$ is substituted by the symmetrized product $\frac{1}{n!}\sum_{\rm permutations} (\hat{p}_{i_1}-A_{i_1}) ... (\hat{p}_{i_n}-A_{i_n})$.
This method of introducing the gauge field to the lattice model differs from the more conventional ways, but it is manifestly gauge invariant, i.e.
the exponent in Eq. (\ref{Z3}) is invariant under the transformation of Eq. (\ref{gt}) if we transform the gauge field as
\begin{equation}
{\bf A}({\bf r}) \rightarrow {\bf A}({\bf r}) + \partial_{{\bf r}} \alpha({\bf r})
\end{equation}
where $\alpha$ is the extension to the continuous values of $\bf r$ of the function of Eq. (\ref{gt}) defined on the discrete lattice points ${\bf r}_n$. Besides, the partition function of Eq. (\ref{Z3}) is obviously reduced to the conventional continuum partition function with the minimal connection of fermion field with the gauge field in the naive continuum limit. Thus, the proposed way of the incorporation of the external gauge field into the lattice theory satisfies all requirements to be fulfilled by the introduction of the gauge field in lattice regularization of quantum field theory.

In case of the models of the solid state physics the proposed introduction of the external gauge field deforms the theory at the interatomic distances. However, the gauge invariance remains intact, and the long distance behavior of the theory should remain the same. Therefore, we consider this appropriate to incorporate the slow varying gauge field to the solid state lattice models in this way.

Now let us come back to momentum space. One can easily check, that Eq. (\ref{Z3}) may be rewritten as
\begin{eqnarray}
Z &=& \int D\bar{\Psi}D\Psi \, {\rm exp}\Big( -  \int_{\cal M} \frac{d^D {\bf p}}{|{\cal M}|} \bar{\Psi}({\bf p})\hat{\cal Q}(i{\partial}_{\bf p},{\bf p})\Psi({\bf p}) \Big)\label{Z4}
\end{eqnarray}
Here
\begin{equation}
\hat{\cal Q} = {\cal G}^{-1}({\bf p} - {\bf A}(i{\partial}^{}_{\bf p}))\label{calQM}
\end{equation}
while the pseudo - differential operator ${\bf A}(i\partial_{\bf p})$ is defined as follows. First, we represent the original gauge field ${\bf A}({\bf r})$ as a series in powers of coordinates ${\bf r}$. Next, variable ${\bf r}$ is substituted in this expansion by the operator $i\partial_{\bf p}$.

In order to prove Eq. (\ref{Z4}) we
introduce the function ${\cal Q}_{right}$ that is constructed of ${\cal G}^{-1}$ as follows. We represent ${\cal G}^{-1}(-i\partial_{\bf r} - {\bf A}({\bf r}))$ as a series in powers of $-i\partial_{\bf r}$ and $ {\bf A}({\bf r})$ such that in each term  $ {\bf A}({\bf r})$ stand right to $-i\partial_{\bf r}$. For example, we represent $(-i \partial_{\bf r} - A({\bf r}))^2$ as $(-i \partial_{\bf r})^2 - 2(-i \partial_{\bf r}) {\bf A}({\bf r}) + {\bf A}^2({\bf r}) - i (\partial {\bf A})$. Next, we substitute the argument of ${\bf A}$ by $i \partial_{\bf p}$ and $-i\partial_{\bf r}$ by $\bf p$. Correspondingly,  ${\cal Q}_{left}$ is defined with the inverse ordering.
With these definitions one may easily prove, that Eq. (\ref{Z3}) is equivalent to
\begin{eqnarray}
Z &=& \int D\bar{\Psi}D\Psi \, {\rm exp}\Big( - \frac{1}{2} \int_{\cal M} \frac{d^D {\bf p}}{|{\cal M}|} \Big[\bar{\Psi}({\bf p}){\cal Q}_{right}(i{\partial}_{\bf p},{\bf p})\Psi({\bf p}) \nonumber\\ &&+ \bar{\Psi}({\bf p}){\cal Q}_{left}(i{\partial}_{\bf p},{\bf p})\Psi({\bf p})  \Big]\Big)\label{Z40}
\end{eqnarray}
after the substitution of Eq. (\ref{Psip}).
Since the commutators $[-i \partial_{r^i},r^j] = i \delta_i^j$ and $[p_i,i \partial_{p_j}] = i \delta_i^j$ are equal to each other, the actual expression for $\frac{1}{2} \Big[{\cal Q}_{right}( i{\partial}^{}_{\bf p},{\bf p}) + {\cal Q}_{left}( i{\partial}^{}_{\bf p},{\bf p}) \Big]$ is given by Eq. (\ref{calQM}).
Finally, the Green function of our system in momentum space satisfies equation
\begin{equation}
\hat{\cal Q}(i \partial_{{\bf p}_1},{\bf p}_1)G({\bf p}_1,{\bf p}_2) = |{\cal M}| \delta^{(D)}({\bf p}_1-{\bf p}_2)\label{QGl}
\end{equation}

\section{Wigner transform in momentum space}
\label{SectWignLat}

According to the proposed above way to incorporate gauge field into the lattice model the Green function in momentum space appears as a correlator
\begin{eqnarray}
G({\bf p}_1,{\bf p}_2)&=&\cor{-} \frac{1}{Z}\int D\bar{\Psi}D\Psi \,\bar{\Psi}({\bf p}_2)\Psi({\bf p}_1)\\ &&{\rm exp}\Big( - \int \frac{d^D {\bf p}}{|{\cal M}|} \bar{\Psi}({\bf p})\hat{Q}(i\partial_{\bf p},{\bf p})\Psi({\bf p}) \Big)\nonumber
\end{eqnarray}
It  obeys equation Eq. (\ref{QGl}).
Wigner transform \cite{Wigner,star,star2,Weyl,berezin} of the Green function may be defined as
\begin{equation}
 \tilde{G}({\bf R},{\bf p}) = \int \frac{d^D{\bf P}}{|{\cal M}|} e^{i {\bf P} {\bf R}} G({\bf p}+{\bf P}/2,{\bf p}-{\bf P}/2)\label{Wl}
\end{equation}
In terms of the Green function in coordinate space this Green function is expressed as:
\begin{equation}
 \tilde{G}({\bf R},{\bf p}) = \sum_{{\bf r}={\bf r}_n} e^{-i {\bf p} {\bf r}} G({\bf R}+{\bf r}/2,{\bf R}-{\bf r}/2)\label{Wl2}
\end{equation}
where
\begin{eqnarray}
G({\bf r}_1,{\bf r}_2)&=& \cor{-}\frac{1}{Z}\int D\bar{\Psi}D\Psi \,\bar{\Psi}({\bf r}_2)\Psi({\bf r}_1)\\ &&{\rm exp}\Big(-\frac{1}{2}\sum_{{\bf r}_n}\Big[ \bar{\Psi}({\bf r}_n)\Big[{\cal G}^{-1}(-i\partial_{\bf r}  \nonumber\\&& - {\bf A}({\bf r}))\Psi({\bf r})\Big]_{{\bf r}={\bf r}_n} + (h.c.)\Big]\Big)\nonumber
\end{eqnarray}

In Appendix A we prove, that this Green function obeys the Groenewold equation (see also Appendix B of \cite{Weyl}):
\begin{eqnarray}
1 &=& {\cal Q}({\bf R},{\bf p})*\tilde G({\bf R},{\bf p})\nonumber\\ && \equiv {\cal Q}({\bf R},{\bf p})e^{\frac{i}{2}(\overleftarrow{\partial}_{\bf R}\overrightarrow{\partial}_{\bf p} - \overleftarrow{\partial}_{\bf p}\overrightarrow{\partial}_{\bf R})}\tilde G({\bf R},{\bf p})   \label{id}
\end{eqnarray}
As well as in continuum case the Weyl symbol \cite{Weyl,berezin} of operator $\hat{\cal Q}$ is given by function $\cal Q$ that depends on the real numbers rather than on the operators. As it is explained in Appendix A, if $\hat{\cal Q}$ has the form of a function ${\cal G}^{-1}$ of the combination $({\bf p} - {\bf A}(\hat{\bf r}))$ with a gauge potential ${\bf A}(\hat{\bf r})$, i.e.
\begin{equation}
\hat{\cal Q}({\bf r},\hat{\bf p}) = {\cal G}^{-1}({\bf p} - {\bf A}(i\partial_{\bf p}))\label{Q}
\end{equation}
then we have
\begin{equation}
{\cal Q}({\bf r},{\bf p}) = {\cal G}^{-1}({\bf p} - {\bf A}({\bf r})) + O([\partial_i A_j]^2)\label{QF}
\end{equation}
Here $O([\partial_i A_j]^2)$ does not contain terms independent of the derivatives of $\bf A$ and the terms linear in those derivatives, i.e. it is higher order in derivatives. In certain particular cases the restrictions on the term $O(([\partial_i A_j]^2)$ may be more strong, or it may even vanish at all \cite{berezin}. In particular, from Sect. 4 (Eqs. (1.23)-(1.27)) of \cite{berezin} it follows, that $O([\partial_i A_j]^2)$ vanishes for the case of Wilson fermions with the Green function $\cal G$ given by Eqs. (\ref{G10}) and (\ref{gWilson}).

Notice, that the star product entering Eq. (\ref{id}) is widely used in deformation quantisation \cite{berezin,star2} and also in some other applications (see, for example, \cite{star} and references therein).

\section{Linear response of electric current to the strength of external gauge field.}
\label{SectLinLat}

Let us apply the gradient expansion to the Wigner transform of the Green function. For this we expand exponent in powers of its arguments in Eq. (\ref{id}). This gives
\begin{eqnarray}
\tilde G({\bf R},{\bf p})  &= &\tilde G^{(0)}({\bf R},{\bf p}) + \tilde G^{(1)}({\bf R},{\bf p}) + ...  \label{Gexp}\\
\tilde G^{(1)}  &= &\cor{+}\frac{i}{2} \tilde G^{(0)} \frac{\partial \Big[\tilde G^{(0)}\Big]^{-1}}{\partial p_i} \tilde G^{(0)}  \frac{\partial  \Big[\tilde G^{(0)}\Big]^{-1}}{\partial p_j} \tilde G^{(0)}
A_{ij} ({\bf R})\nonumber
\end{eqnarray}
Here $\tilde G^{(0)}({\bf R},{\bf p})$ is defined as the Green function with the field strength $A_{ij} = \partial_i A_j - \partial_j A_i$ neglected. It is given by
\begin{eqnarray}
&&\tilde G^{(0)}({\bf R},{\bf p})  = {\cal G}({\bf p}-{\bf A}({\bf R}))\label{Q0}
\end{eqnarray}

Next, suppose, that we modified the external gauge field as ${\bf A} \rightarrow {\bf A} + \delta {\bf A}$. The response to this extra contribution to gauge potential gives electric current. Let us calculate this response basing on the description of the system given by Eq. (\ref{Z4}):
\begin{eqnarray}
{\delta} \, {\rm log}\, Z&=& -\frac{1}{Z} \int D\bar{\Psi}D\Psi \, {\rm exp}\Big( - \int_{\cal M} \frac{d^D {\bf p}}{|{\cal M}|} \bar{\Psi}({\bf p})\hat{\cal Q}(i{\partial}_{\bf p},{\bf p})\Psi({\bf p}) \Big) \, \int_{\cal M} \frac{d^D {\bf p}}{|{\cal M}|} \bar{\Psi}({\bf p})\Big[\delta \hat{\cal Q}(i{\partial}_{\bf p},{\bf p})\Big]\Psi({\bf p}) \nonumber\\ & = & \cor{+}\int_{\cal M} \frac{d^D {\bf p}}{|{\cal M}|} \, {\rm Tr} \, \Big[ \delta \hat{\cal Q}(i{\partial}_{{\bf p}_1},{\bf p}_1)\Big]G({\bf p}_1,{\bf p}_2)\Big|_{{\bf p}_1 = {\bf p}_2 = {\bf p}}\nonumber\\ & = & \cor{+}\sum_{{\bf R}={\bf R}_n}\int_{\cal M} \frac{d^D {\bf p}}{|{\cal M}|} \,  {\rm Tr} \, \Big[ \delta \hat{\cal Q}(i{\partial}_{{\bf P}}+i{\partial}_{{\bf p}}/2 ,{\bf p}+{\bf P}/2)\Big]\, e^{-i {\bf P}{\bf R}} \tilde{G}({\bf R},{\bf p})\Big|_{{\bf P} = 0}  \label{j4}
\end{eqnarray}
In Appendix B we discuss the Weyl symbol ${\cal Q}({\bf r},{\bf p})$ of the operator $\hat{\cal Q}$ entering Eq. (\ref{id}). Notice, that $2{\bf p}$ and ${\bf P}$ enter the expression inside $\hat{\cal Q}$ in a symmetric way. This allows to use Eq. (\ref{corrl}). The form of Eq. (\ref{j4}) demonstrates, that the above expression for the electric current may also be written through the function $\cal Q$:
\begin{eqnarray}
{\delta} \, {\rm log}\, Z &=& \cor{+}\sum_{{\bf R}={\bf R}_n}\int_{\cal M} \frac{d^D {\bf p}}{|{\cal M}|} \,  {\rm Tr} \, \Big[ \delta {\cal Q}(i\overrightarrow{\partial}_{{\bf P}}-i\overleftarrow{\partial}_{{\bf p}}/2 ,{\bf p}+{\bf P}/2)\Big]\, \nonumber\\&&e^{-i {\bf P}{\bf R}} \tilde{G}({\bf R},{\bf p})\Big|_{{\bf P} = 0}\nonumber\\&=& \cor{+}\sum_{{\bf R}={\bf R}_n}\int_{\cal M} \frac{d^D {\bf p}}{|{\cal M}|} \,  {\rm Tr} \, \Big[ \delta {\cal Q}({\bf R} ,{\bf p}+{\bf P}/2)\Big]\, \nonumber\\&&e^{-i {\bf P}{\bf R}} \tilde{G}({\bf R},{\bf p})\Big|_{{\bf P} = 0}   \label{j428}
\end{eqnarray}
According to the notations of Appendix B the arrows above the derivatives mean, that those derivatives act only outside of ${\cal Q}$, and do not act on the arguments of $\cal Q$, i.e. $\overleftarrow\partial_{\bf p}$ acts on the function equal to $1$ standing left to the function $\cal Q$ while $\overrightarrow\partial_{\bf P}$ acts on the exponent $e^{-i {\bf P}{\bf R}} $.

As a result of the above mentioned manipulations we come to the following simple expression for the electric current per unit volume of coordinate space, which follows from the relation $\delta \, {\rm log}\, Z = \sum_{{\bf R}={\bf R}_n}{j}^k({\bf R}) \delta A_k({\bf R})|{\cal V}|$:
\begin{eqnarray}
j^k({\bf R}) &=& \cor{-}\int_{\cal M} \frac{d^D {\bf p}}{|{\cal V}||{\cal M}|} \,  {\rm Tr} \, \tilde{G}({\bf R},{\bf p}) \frac{\partial}{\partial p_k}\Big[\tilde{G}^{(0)}({\bf R},{\bf p})\Big]^{-1}\label{j423}
\end{eqnarray}
Here by $|{\cal V}|$ we denote the volume of the unit cell understood as the ratio of the total volume of the system to the number of lattice points at which the field $\Psi$ is defined. For the ordinary hypercubic lattice the product of the two volumes is obviously  equal to $(2\pi)^D$. One might think, that for the lattices of more complicated symmetry the product of the momentum space volume and the defined above volume of the lattice cell may differ from this expression. Nevertheless, this is not so, and in general case the given product is always equal to $(2 \pi)^D$ exactly. In general case of an arbitrary crystal the direct proof is rather complicated. However, the result for the product of the two volumes may be found from the simple field theoretical correspondence: the limit of the microscopic model described by the effective low energy theory should correspond to the product of the two volumes equal to $(2\pi)^D$. Notice, that the construction of the unit cell in the original lattice should be performed with care. One has to count only those sites of the original crystal lattice, at which the dynamical variables of the model described by Eq. (\ref{Z3}) are incident.

Let us apply the gradient expansion to Eq. (\ref{j423}). It results in the following expression for the electric current:
\begin{eqnarray}
j^{k}({\bf R}) &=& j^{(0)k}({\bf R}) + j^{(1)k}({\bf R}) + ...\nonumber\\
j^{(0)k}({\bf R})  & = &\cor{-} \int \frac{d^D {\bf p}}{(2\pi)^D}\,  {\rm Tr}\, \tilde G^{(0)}({\bf R},{\bf p})\frac{\partial  \Big[\tilde G^{(0)}({\bf R},{\bf p})\Big]^{-1}}{\partial p_k}\label{j142}
\end{eqnarray}
Notice, that the second row of this expression represents the topological invariant as long as we deal with the system with regular Green functions, which do not have poles or zeros, i.e. this expression is unchanged while we are continuously deforming the Green function. We will not need this expression below since it does not contain the linear response to the external field strength.

In the $3+1$ D systems the contribution to this current originated from $\tilde{G}^{(1)}$ is given by
\begin{eqnarray}
j^{(1)k}({\bf R})  &= &\cor{-} \frac{1}{4\pi^2}\epsilon^{ijkl} {\cal M}_{l} A_{ij} ({\bf R}), \label{calM}\\
{\cal M}_l &=& \int_{} \,{\rm Tr}\, \nu_{l} \,d^4p \label{Ml} \\ \nu_{l} & = &  - \frac{i}{3!\,8\pi^2}\,\epsilon_{ijkl}\, \Big[  {\cal G} \frac{\partial {\cal G}^{-1}}{\partial p_i} \frac{\partial  {\cal G}}{\partial p_j} \frac{\partial  {\cal G}^{-1}}{\partial p_k} \Big]  \label{nuG}
\end{eqnarray}

In the linear response theory we should substitute ${\bf A}=0$ into the expression for ${\cal M}_l$. Therefore, in Eq. (\ref{nuG}) we substitute $\cal G$ instead of $\tilde{G}^{(0)}$.  ${\cal M}_l$ is the topological invariant, i.e. it is robust to any variations of the Green function $\tilde{G}$ as long as the singularities are not encountered (for the proof see Appendix B).

Although Eq. (\ref{calM}) was derived for the case of compact momentum space, the non - compact case may always be obtained as a limit of the compact case, when a certain parameter of the theory tends to infinity. Therefore, in Eq. (\ref{calM}) the integration may be over compact or over non - compact space depending on the type of the given system. In particular, we will encounter in the upcoming sections the non - interacting condensed matter systems with the Green functions of the form ${\cal G}^{-1} = i \omega - \hat{H}$, where $\omega \in R$ is the imaginary frequency while $\hat{H}$ is the Hamiltonian. Momentum space of such a system is non - compact and has the form of $R\otimes \Omega$, where $\Omega$ is the compact Brillouin zone. The linear response of electric current to external electromagnetic field is still given in such systems by Eq. (\ref{calM}). The derivation is analogous to the one we gave above for the case of compact momentum space.

\section{ $2+1$ D anomalous quantum Hall effect}
\label{SectHall}

The considerations of the previous section may easily be applied to the $2+1$ D systems. Then instead of Eq. (\ref{calM}) we arrive at
\begin{eqnarray}
j^{(1)k}({\bf R})  &= & \cor{-} \frac{1}{4\pi}\epsilon^{ijk} {\cal M}_{} A_{ij} ({\bf R}),\quad  {\cal M} = \int_{} \,{\rm Tr}\, \nu_{} \,d^3p \label{j2d}\\ \nu_{} & = &  - \frac{i}{3!\,4\pi^2}\,\epsilon_{ijk}\, \Big[  \tilde G^{(0)}({\bf R},{\bf p}) \frac{\partial \Big[\tilde G^{(0)}({\bf R},{\bf p})\Big]^{-1}}{\partial p_i} \frac{\partial  \Big[\tilde G^{(0)}({\bf R},{\bf p})\Big]}{\partial p_j} \frac{\partial  \Big[\tilde G^{(0)}({\bf R},{\bf p})\Big]^{-1}}{\partial p_k} \Big]  \label{calM2d}
\end{eqnarray}
In this section we give the derivation of the well - known expression for the Hall current in the gapped system through the topological invariant in momentum space. For the references to the other derivations see, for example, \cite{Volovik2003}.

Let us consider the $2+1 $ D model with the gapped fermions and the Green function ${\cal G}({\bf p})$ that depends on the three - vector ${\bf p} = (p_1,p_2,p_3)$ of Euclidean momentum. In order to obtain expression for the Hall current let us introduce into  Eq. (\ref{calM2d}) the external electric field ${\bf E} = (E_1,E_2)$ as $A_{3k} = -i E_k$ (the third component of vector corresponds to imaginary time). This results in the following expression for the Hall current
\begin{equation}
{j}^k_{Hall} = \cor{-}\frac{1}{2\pi}\,\tilde{\cal N}_3\,\epsilon^{ki}E_i,\label{HALLj}
\end{equation}
where the topological invariant (denoted by $\tilde{\cal N}_3$ according to the classification of \cite{Volovik2003}) is to be calculated for the original system with vanishing background gauge field:
\begin{eqnarray}
\tilde{\cal N}_3 &=&  -\frac{1}{24 \pi^2} {\rm Tr}\, \int_{} {\cal G}^{-1} d {\cal G} \wedge d {\cal G}^{-1} \wedge d {\cal G}\label{N3A}
\end{eqnarray}
Eq. (\ref{N3A}) defines the topological invariant (this is proved in Appendix B). Its appearance explains the quantization of Hall conductance. In Appendix C we demonstrate how to calculate the invariant $\tilde{\cal N}_3$ entering Eq. (\ref{HALLj}) for the models with $2\times2$ Green functions, and  give the example of the model, where the corresponding value is nonzero. (For the other examples of such models see \cite{Volovik2003}.) In the considered example the Green function has the form ${\cal G}^{-1} = i \omega - H({\bf p})$, where the Hamiltonian is
\begin{equation}
H = {\rm sin}\,p_1\, \sigma^2 - {\rm sin}\, p_2 \, \sigma^1 - (m + \sum_{i=1,2}(1-{\rm cos}\,p_i)) \, \sigma^3\label{Ham0}
\end{equation}
For $m \in (-2,0)$ we have
$\tilde{\cal N}_3 = 1$ while  $\tilde{\cal N}_3  = -1 $ for $m\in (-4,-2)$ and $\tilde{\cal N}_3  = 0 $ for $m\in (-\infty,-4)\cup (0,\infty)$.


Let us consider briefly  the above mentioned system with the Hamiltonian Eq. (\ref{Ham0}) in the presence of boundary. It gives rise to the potential step of electric potential and the step in $m$, which becomes the function of ${\bf R}$. Say, if the boundary is directed along the $x$ axis, the potential step results in the extra electric field directed towards the interior of the sample
\begin{equation}
E_2 = \Delta \phi \delta(y),
\end{equation}
where $\Delta \phi$ is the jump of electric potential. We may suppose, that
\begin{eqnarray}
m &= & m^{(0)} \in (-2,0), \quad   y>0\nonumber\\
m &= & m^{(-)} \in (0,\infty), \quad y<0
\end{eqnarray}
Then at $y>0$ we deal with the AQHE system with $\tilde{\cal N}_3=1$ while at $y<0$ there is the ordinary insulator with $\tilde{\cal N}_3=0$. The step in $m({\bf R})$ is important for the demonstration of the appearance of massless edge states. Namely, at $m({\bf R}) = 0$ the Green function $\tilde{G}^{(0)}$ has the pole, which corresponds to the massless excitation. Such excitations have definite chirality, and in general case their number is proportional to the jump of the topological invariant $\tilde{\cal N}_3$ across the boundary (if we substitute into the expression for $\tilde{\cal N}_3$ the $\bf R$ - dependent function $m$). For the derivation of this index theorem the reader is advised to consult chapter 22 of \cite{Volovik2003}. The similar derivation for the $3+1$ D case will be given in Sect. \ref{bulkboundary}.

\section{Topological invariant $\tilde{\cal N}_3$ and Berry curvature}
\label{SectBerry}

In this section we demonstrate how the topological invariant $\tilde{\cal N}_3$ is expressed through Berry curvature in the $2+1$ D models. This correspondence works for the Green function of the non - interacting model that has the form
\begin{equation}
{\cal G}^{-1} = i\omega - \hat{H}
\end{equation}
with the hermitian hamiltonian $\hat{H}$. In the presence of interactions, when the Green function receives a more complicated form this correspondence looses its sense. (This is the advantage of the formalism, that utilizes the Green functions.) Nevertheless, at the present moment the approach to the description of QHE based on Berry curvature is more popular. Therefore, we feel this instructive to derive it from our expressions.

Thus, we start from
\begin{eqnarray}
\tilde{\cal N}_3 &=&  \frac{1}{24 \pi^2} {\rm Tr}\, \int_{} {\cal G} d {\cal G}^{-1} \wedge d {\cal G} \wedge d {\cal G}^{-1}\label{N3A2}
\end{eqnarray}
and substitute into it
\begin{equation}
{\cal G}^{-1} = i\omega - \sum_n{\cal E}_n(\vec{p})|n,\vec{p}\rangle\langle n,\vec{p}|
\end{equation}
Here ${\cal E}_n({\bf p})$ is the $n$ - th eigenvalue of the Hamiltonian depending on to the lattice momentum $\vec{p} = (p_1,p_2)$. The corresponding eigenvector of the Hamiltonian is denoted by $|n,\vec{p}\rangle$. We have:
\begin{eqnarray}
\tilde{\cal N}_3 &=&  \frac{i\epsilon^{ij}}{8 \pi^2}\sum_{nk}\, \int \frac{d\omega\,d^2p\, \langle n,\vec{p}|\partial_i \hat{H}(\vec{p}) |k,\vec{p}\rangle \langle k,\vec{p}|\partial_j\hat{H}(\vec{p}) |n,\vec{p}\rangle}{(i\omega-{\cal E}_n(\vec{p}))^2(i\omega-{\cal E}_k(\vec{p}))}
\end{eqnarray}
Integral over the imaginary frequency $\omega$ may be taken as a residue at the negative value of ${\cal E}_k$, and gives
\begin{eqnarray}
\tilde{\cal N}_3 &=&  \frac{i\epsilon^{ij}}{2\pi}\sum_{n\ne k}\, \int \frac{d^2p\, \langle n,\vec{p}|\partial_i \hat{H}(\vec{p}) |k,\vec{p}\rangle \langle k,\vec{p}|\partial_j\hat{H}(\vec{p}) |n,\vec{p}\rangle}{({\cal E}_k(\vec{p})-{\cal E}_n(\vec{p}))^2}\theta(-{\cal E}_k(\vec{p}))
\end{eqnarray}
This expression is the starting point of \cite{TTKN}, where the expression for the Hall conductivity through the Berry curvature has been derived for the first time. After some algebra we arrive at
\begin{eqnarray}
\tilde{\cal N}_3 &=&  \frac{i\epsilon^{ij}}{2 \pi}\sum_{k}\, \int d^2p\, \partial_i\langle k,\vec{p}|\partial_j |k,\vec{p}\rangle \theta(-{\cal E}_k(\vec{p}))
\end{eqnarray}
With the following definition of the momentum space gauge field
\begin{equation}
{\cal A}_j = i \langle k,\vec{p}|\partial_j |k,\vec{p}\rangle \label{AC}
\end{equation}
we arrive at
\begin{eqnarray}
\tilde{\cal N}_3 &=&  \frac{\epsilon^{ij}}{4\pi}\sum_{k:{\cal E}_k < 0}\, \int d^2p\, {\cal F}_{ij}
\end{eqnarray}
where the sum is over the occupied states while Berry curvature is given by
\begin{eqnarray}
{\cal F}_{ij} &=&   \partial_i {\cal A}_j - \partial_j {\cal A}_i \label{BC}
\end{eqnarray}

\section{AQHE in the $3+1$D systems}
\label{SectHall3d}

\subsection{General case}

Now let us turn to the $3+1$D systems. Similar to the two - dimensional models the Hall current is given by
\begin{equation}
{j}^k_{Hall} =\cor{-} \frac{1}{4\pi^2}\,{\cal M}^\prime_l\,\epsilon^{jkl}E_j,\label{HALLj3d}
\end{equation}
where ${\cal M}^\prime_l = i{\cal M}_l/2$ is given by
\begin{eqnarray}
{\cal M}^\prime_l &=&  \frac{1}{3!\,4\pi^2}\,\epsilon_{ijkl}\,\int_{} \,\,d^4p\,{\rm Tr} \Big[  {\cal G} \frac{\partial {\cal G}^{-1}}{\partial p_i} \frac{\partial  {\cal G}}{\partial p_j} \frac{\partial  {\cal G}^{-1}}{\partial p_k} \Big]  \label{nuGHall}
\end{eqnarray}

Notice, that in the case of the non - interacting condensed matter system with the Green function  ${\cal G}^{-1} = i \omega - \hat{H}({\bf p})$ expressed through the Hamiltonian $\hat{H}$ we may derive (similar to the above considered case of the $2+1$ D system) the representation of the components of topological invariant ${\cal M}^{\prime}_l$ with $l\ne 4$ through the Berry curvature given by Eqs. (\ref{AC}), (\ref{BC}):
\begin{eqnarray}
{\cal M}^\prime_l &=&  \frac{\epsilon^{ijl}}{4\pi}\sum_{\rm occupied}\, \int d^3p\, {\cal F}_{ij}
\end{eqnarray}
Here the sum is over the occupied branches of spectrum.

\subsection{The case of $2\times 2$ Green functions}
\label{Sect22}

Let us demonstrate how ${\cal M}^\prime_l$ may be calculated. Let us consider first (as in Appendix C) the case, when the Green function has the form
 \begin{equation}
 {\cal G}^{-1}({\bf p}) = i\sigma^3\Big(\sum_{k}\sigma^{k} g_{k}({\bf p}) - i g_4({\bf p})\Big)\label{G12d2}
 \end{equation}
Now we have ${\bf p}=(p^1,p^2,p^3,p^4)$, and $p^4$ is to be identified with the imaginary frequency.
We  denote $\hat{g}_k = \frac{g_k}{g}$, $g = \sqrt{\sum_{k=1,2,3,4}g_k^2}$, and  introduce the parametrization
\begin{equation}
\hat{g}_4 = {\rm sin}\,\alpha, \quad \hat{g}_a = k_a\,{\rm cos}\,\alpha
\end{equation}
where $a=1,2,3$ while $\sum_{a=1,2,3}k^2_a=1$, and $\alpha \in [-\pi/2,\pi/2]$. We suppose, that $\hat{g}_4({\bf p})=0$ on the boundary of momentum space ${\bf p}\in \partial {\cal M}$. This gives
  \begin{eqnarray}
{\cal M}^\prime_n &=&  \frac{1}{4 \pi^2} \epsilon^{abc}\,\epsilon^{ijkn} \int_{\cal M} \, {\rm cos}^2 \alpha\, k_a\,  \partial_i\alpha \, \partial_j k_b\, \partial_k k_c\, d^4p \nonumber\\ &=&  \frac{1}{4 \pi^2} \epsilon^{abc}\, \int_{\cal M} \, k_a\,  d(\alpha/2+\frac{1}{4}{\rm sin}\,2\alpha)  \wedge d k_b \wedge d k_c\wedge dp_n \nonumber\\ &=& -\sum_l \frac{1}{4 \pi^2} \epsilon^{abc}\, \int_{\partial{\Omega(y_l)}} \, k_a\,  (\alpha/2+\frac{1}{4}{\rm sin}\,2\alpha)   d k_b \wedge d k_c \wedge dp_n
\end{eqnarray}
Now $\Omega(y_l)$ is the small vicinity of line $y_l(s)$ in momentum space, where vector $k_i$ is undefined. Along these lines $\alpha \rightarrow \pm\pi/2$.
We have
  \begin{eqnarray}
{\cal M}^\prime_j &=&   -\frac{1}{2}\sum_l \, \int_{y_l(s)} \, {\rm sign}(g_4(y_l)) \,{\rm Res}\,(y_l) dp_j\label{calMproj}
\end{eqnarray}
Here we use the notations of \cite{VZ2012}, and
\begin{eqnarray}
{\rm Res}\,(y_l) &=&  \frac{1}{8 \pi} \epsilon^{ijk}\, \int_{\Sigma(y_l)} \, \hat{g}_i  d \hat{g}_j \wedge d \hat{g}_k
\end{eqnarray}
is the integer number, the corresponding integral is  along the  infinitely small surface $\Sigma$, which is wrapped around the line $y_j(s)$ near to the given point of this line.

\subsection{The case of $4\times 4$ Green functions}

Let us consider the more complicated systems with the Green function of the form
 \begin{equation}
 {\cal G}^{}({\bf p}) = \Big(\sum_{k}\gamma^{k} g_{k}({\bf p})+ \gamma^5 g_5({\bf p}) + \gamma^3 \gamma^5 b({\bf p})\Big)^{-1}\label{G202}
 \end{equation}
 where $\gamma^k$ are Euclidean Dirac matrices while $g_k({\bf p})$, $b({\bf p})$ are the real - valued functions, $k = 1,2,3,4$.
We define $\gamma^5$  in chiral representation as ${\rm diag}(1,1,-1,-1)$.
Let us represent ${\cal G}^{-1}$ as follows:
 \begin{equation}
 {\cal G}^{-1}({\bf p}) = e^{\alpha \gamma^3\gamma^5}\Big(\gamma^{1} g_{1}({\bf p})+\gamma^{2} g_{2}({\bf p})+\gamma^{4} g_{4}({\bf p}) +\gamma^{3} \sqrt{g_3^2({\bf p})+g^2_5({\bf p})}   + \gamma^3 \gamma^5 b({\bf p})\Big) e^{-\alpha \gamma^3\gamma^5}\label{G2031}
 \end{equation}
where
\begin{eqnarray}
\alpha &=& -\frac{1}{2}{\rm arctg}\,\frac{g_5}{g_3}
\end{eqnarray}
Let us assume, that $g_3$ never equals to zero, and that the Green function does not have poles. Then since ${\cal M}^\prime_l$ is the topological invariant, we may deform continuously the Green function to the form
 \begin{equation}
 {\cal G}({\bf p}) = \Big(\gamma^{1} g_{1}({\bf p})+\gamma^{2} g_{2}({\bf p})+\gamma^{4} g_{4}({\bf p}) +\gamma^{3} \sqrt{g_3^2({\bf p})+g^2_5({\bf p})}   + \gamma^3 \gamma^5 b({\bf p})\Big)^{-1}\label{G20312}
 \end{equation}
 Under this deformation ${\cal M}^\prime_l$ remains invariant if on the boundary of momentum space
 \begin{equation}{\rm Tr} \left(
[\delta {\cal G}^{-1}] {\cal G}]
  d  {\cal G}^{-1}\wedge
  d  {\cal G}\right) = 0\label{condtop0}\end{equation}
  This follows from Appendix B. This condition is fulfilled if on $\partial {\cal M}$
  \begin{equation}{\rm Tr} \left(
\gamma^{A} {\cal G}
  d  {\cal G}^{-1}\wedge
  d  {\cal G}\right) = 0, \quad A = 3,5 \label{cond00}\end{equation}

Now ${\cal M}^\prime_l$ is equal to
\begin{eqnarray}
{\cal M}^\prime_l &=& {\cal M}^\prime_{l,+} +  {\cal M}^\prime_{l,-}\label{Mpm}
\end{eqnarray}
where ${\cal M}^\prime_{l,\pm}$ is to be calculated using the $2\times 2$ Green functions
 \begin{eqnarray}
 &&{\cal G}^{-1}({\bf p}) = \sigma^3 \Big(\sum_{k}\sigma^{k} g^\prime_{k}({\bf p}) - i g^\prime_4({\bf p})\Big)\nonumber\\&&
g^\prime_1 = g_{2}({\bf p}), \quad g^\prime_2 = -g_{1}({\bf p}), \quad g^\prime_3 = g_{4}({\bf p}), \quad g^\prime_4 = \mp  \Big(\sqrt{g_3^2({\bf p})+g_5^2({\bf p})} \pm b({\bf p})\Big)  \label{G12d23}
 \end{eqnarray}
Thus the problem is reduced to the one considered above in Sect. \ref{Sect22}. It is worth mentioning, that the given way of calculation may easily be extended to the Green functions of some other forms. In particular, the same expression for the topological invariant is valid for ${\cal G}^{}({\bf p}) = \Big(\sum_{k}\gamma^{k} g_{k}({\bf p}) - i  g_5({\bf p}) + \gamma^3 \gamma^5 b({\bf p})\Big)^{-1}$.

\section{Bulk - boundary correspondence}
\label{bulkboundary}

We are able to use the Wigner transform of the Green function in order to describe the bulk - boundary correspondence (for the description of the method see Chapter 22 in \cite{Volovik2003} and \cite{EssinGurarie2011}). Let us suppose, that in the given solid (${\bf R}$ is inside the bulk) there are no poles of $\tilde{G}^{(0)}({\bf R},{\bf p})$ as functions of ${\bf p}$. Let us also suppose, that a surface divides space into the two parts with different values of ${\cal M}^\prime_l$. For simplicity let us consider the situation, when momentum space has the form of torus, ${\cal M}^{\prime}_l$ is nonzero for $l = 1$ only, while this surface is situated in the $xy$ plane, and $\tilde{G}^{(0)}({\bf R},{\bf p})$ depends on $R_3 = z$ only. The set $z,p_k$ with $k\ne l$ constitues four - dimensional space. The following integrals over the two hyper - surfaces in this space with $z = \pm \epsilon$ (with small $\epsilon$) may be deformed safely to a closed hyper - surface:
\begin{eqnarray}
{\cal N}_3(p_1)&=&  \frac{1}{3!\,4\pi^2}\,\epsilon_{ijkl}\,\int_{} \,\,dp_2 dp_3 dp_4\,{\rm Tr} \Big[  \tilde{G}^{(0)} \frac{\partial (\tilde{G}^{(0)})^{-1}}{\partial p_i} \frac{\partial  \tilde{G}^{(0)}}{\partial p_j} \frac{\partial  (\tilde{G}^{(0)})^{-1}}{\partial p_k} \Big]  \label{nuGHall21}
\end{eqnarray}
If the resulting integral is nonzero, then inside it there are ${\cal N}_3$ topologically protected poles of $\tilde{G}^{(0)}$, which correspond to massless states concentrated in a small vicinity of boundary.
There is also the question about the relation of these poles of $\tilde{G}^{(0)}$ with the gapless excitations in the effective theory that describes fermions on the boundary. Such a relation was discussed in Chapter 22 of \cite{Volovik2003} and in \cite{EssinGurarie2011}). The output of this study is that the chirality of those $1+1$ dimensional fermions is given by the sign of ${\cal N}_3$. This chirality is encoded in the quantity
\begin{equation}
{\cal N}_1({\cal C},p_1) = \frac{1}{2\pi i}\, \sum_z \,\int \frac{dp_3}{2\pi} \int_{{\cal C}} {\rm Tr}\, \tilde{G}({\bf R},{\bf p}) d \tilde{G}^{-1}({\bf R},{\bf p})\label{N1}
\end{equation}
Here the contour is in the plane $(p_4, p_2)$, while the sum is over all possible (discrete) values of $z$.
Again, we use Eq. (\ref{id}) and expand exponent in powers of its arguments, which gives
\begin{eqnarray}
\tilde G({\bf R},{\bf p})  &= &\tilde G^{(0)}({\bf R},{\bf p}) + \tilde G^{(1)}({\bf R},{\bf p}) + ...  \label{Gexp2}\\
\tilde G^{(1)}  &= &\frac{i}{2} \tilde G^{(0)} \frac{\partial \Big[\tilde G^{(0)}\Big]^{-1}}{\partial p_3} \tilde G^{(0)}  \frac{\partial  \Big[\tilde G^{(0)}\Big]^{-1}}{\partial z} \tilde G^{(0)}
\nonumber\\
 & &-\frac{i}{2} \tilde G^{(0)} \frac{\partial \Big[\tilde G^{(0)}\Big]^{-1}}{\partial z} \tilde G^{(0)}  \frac{\partial  \Big[\tilde G^{(0)}\Big]^{-1}}{\partial p_3} \tilde G^{(0)}
\nonumber
\end{eqnarray}
Here $\tilde G^{(0)}({\bf R},{\bf p})$ is given by Eq. (\ref{Q0}). Let us substitute Eq. (\ref{Gexp2}) into Eq. (\ref{N1}). Assuming, that $\tilde{G}$ is slowly varying we substitute $\sum_z$ by $\int_{-\infty}^{\infty} dz$. This way we come to the integral of the form
\begin{eqnarray}
{\cal N}_3(p_1)&=&  \frac{1}{3!\,4\pi^2}\,\,\int_{} \,{\rm Tr} \Big[  \tilde{G}^{(0)} d (\tilde{G}^{(0)})^{-1}\wedge d  \tilde{G}^{(0)} \wedge d(\tilde{G}^{(0)})^{-1} \Big]  \label{nuGHall22}
\end{eqnarray}
Here the integral is over the surface ${\cal C}\otimes S^1 \otimes R^1$, where $S^1$ corresponds to the values of $p_3$ while $R^1$ corresponds to the values of $z$. Since there are no poles of $\tilde{G}$ inside the bulk we are able to deform this hyper - surface into the form of two hyper - planes $(p_2,p_3,p_4)$ placed at $z = \pm \epsilon$, which gives the value of ${\cal N}_3(p_1)$.

In turn, for the case of noninteracting system with ${\cal G} = i\omega - \hat{H}$ we have
$${\cal G} = i\omega - \sum_n {\cal E}_n(p_1,p_2)|n,p_1,p_2\rangle\langle n,p_1,p_2|$$ where the sum is over the eigenstates of the Hamiltonian enumerated by index $n$ and the values of conserved momenta $p_1,p_2$. Let us denote ${\cal G}_n = i\omega - {\cal E}_n(p_1,p_2)$. Then
Eq. (\ref{N1}) may be represented as
\begin{equation}
{\cal N}_1({\cal C},p_1) = \frac{1}{2\pi i}\, \sum_n \, \int_{{\cal C}} \, \tilde{G}_n d \tilde{G}^{-1}_n \label{N12}
\end{equation}
This expression gives the sum of the chiralities of the states enumerated by $n$. Only those states which are localized at the boundary may have gapless excitations because according to our supposition in the bulk of the system there are no poles of the Green function. Thus we come to conclusion, that ${\cal N}_3(p_1)$ enumerates the gapless boundary modes, and its sign corresponds to their chirality.
Notice, that the similar conclusion is given in the lattice theory of domain - wall fermions (see, for example, \cite{Kaplan1992}).

Now let us describe the dependence of this pattern on $p_1$.
${\cal N}_3(p_1)$ cannot depend on $p_1$ because we assumed, that there are no poles in the bulk. Therefore, we come to the conclusion, that the topologically protected gapless surface states exist, which form lines in momentum space (the dispersion does not depend on $p_1$). The number of such Fermi lines is equal to the jump of ${\cal N}_3$ accross the boundary and is related to the jump of ${\cal M}^\prime_1 = \int dp_1 {\cal N}_3(p_1)$. The generalization of this consideration to the values of ${\cal M}^\prime_l$ for arbitrary $l$ is straightforward. The generalization to the case of the Brillouin zones of arbitrary form is more involved and will be omitted here. It is worth mentioning, that in section \ref{SectWeylSem} we will see, that for the Weyl semimetals, when there are Fermi points in the bulk, the described above boundary Fermi lines become the Fermi arcs connecting the bulk Fermi points.

\section{AQHE in the toy models of $3+1$ D  insulators}
\label{SectIns}

\subsection{Model I1}

In this section we will consider the two particular examples of the $3+1$D insulators, for which the topological invariant ${\cal M}^\prime_l$ may be calculated using the technique developed above. Let us start from the model with the Hamiltonian
\begin{equation}
H = {\rm sin}\,p_1\, \sigma^2 - {\rm sin}\, p_2 \, \sigma^1 - (m^{(0)}-\gamma\, {\rm cos}\,p_3 + \sum_{i=1,2}(1-{\rm cos}\,p_i)) \, \sigma^3
\end{equation}
For $\gamma < 1$, and $m^{(0)} \in (-2+\gamma,-\gamma)$ we deal with the insulator. Here we use the system of units, in which the lattice spacing $a$ is equal to unity. Therefore, lattice momentum $p_k$ is dimensionless. The $2\times 2$ Green function has the form ${\cal G}^{-1} = i \omega - H({\bf p})$ while the Brillouin zone has the form of torus.
We will be using the following expression for the Green function:
\begin{equation}
-i\sigma^3{\cal G}^{-1} = {\rm sin}\,p_1\, \sigma^1 + {\rm sin}\, p_2 \, \sigma^2 +  \omega \, \sigma^3 - i (m^{(0)} - \gamma {\rm cos}\,p_3+ \sum_{i=1,2}(1-{\rm cos}\,p_i))
\end{equation}
We have
$$\hat{g}_4({\bf p}) = \frac{(m^{(0)} - \gamma {\rm cos}\,p_3 + \sum_{i=1,2}(1-{\rm cos}\,p_i))}{\sqrt{(m^{(0)}- \gamma {\rm cos}\,p_3 + \sum_{i=1,2}(1-{\rm cos}\,p_i))^2+ {\rm sin}^2\,p_1+ {\rm sin}^2\,p_2 + \omega^2}} $$
and
\begin{eqnarray}
\hat{g}_4({\bf p}) & = & 0, \quad {\bf p}\in \partial{\cal M} \quad (\omega \rightarrow \pm \infty)\nonumber\\
\hat{g}_4({\bf p}) & = & -1, \quad \hat{g}_i({\bf p})  = 0\quad (k = 1,2,3),\quad  {\bf p} = (0,0,p_3,0),\quad p_3\in (-\pi,\pi)
\nonumber\\
\hat{g}_4({\bf p}) & = & 1,\quad  \hat{g}_i({\bf p})  = 0\quad (k = 1,2,3),\quad  {\bf p} = (0,\pi,p_3,0),\quad p_3\in (-\pi,\pi) \nonumber\\
\hat{g}_4({\bf p}) & = & 1,\quad  \hat{g}_i({\bf p})  = 0\quad (k = 1,2,3),\quad  {\bf p} = (\pi,0,p_3,0),\quad p_3\in (-\pi,\pi)\nonumber\\
\hat{g}_4({\bf p}) & = & 1,\quad  \hat{g}_i({\bf p})  = 0\quad (k = 1,2,3),\quad  {\bf p} = (\pi,\pi,p_3,0),\quad p_3\in (-\pi,\pi)\label{listzeros}
\end{eqnarray}
Therefore ${\cal M}^\prime_1 = {\cal M}^\prime_2 ={\cal M}^\prime_4 = 0$ while
\begin{eqnarray}
{\cal M}^\prime_3 &=&   \frac{2\pi}{2}- \frac{2\pi}{2} (-1)-\frac{2\pi}{2}(-1) - \frac{2\pi}{2} = 2 \pi
\end{eqnarray}
One can see, that we deal with the AQHE current
\begin{equation}
 {j}^k_{Hall} =\cor{-} \frac{1}{2\pi\,a}\,\epsilon^{jk3}E_j
\end{equation}
We restored the value of the lattice spacing $a$ in this expression in order to keep the track of dimensionality.
Notice, that the same expression may be read off from Eq. (A.5) of \cite{Klinkhamer:2004hg}.

Eq. (\ref{nuGHall}) represents the topological invariant on the boundary of momentum space
\begin{equation}{\rm Tr} \left(
\{[\delta {\cal G}^{-1}] {\cal G}] \}
  d  {\cal G}^{-1}\wedge
  d  {\cal G}\right) = 0\label{condtop}\end{equation}
  This follows from Appendix B.
The interactions, that keep Eq. (\ref{condtop}) cannot affect the value of the AQHE current until the phase transition is encountered.

\subsection{Model I2}
Our second example corresponds to the system with the $4\times 4$ Green function of the form of Eq. (\ref{G202}) with
\begin{eqnarray}
&& g_1({\bf p}) = -{\rm sin}\,
p_2, \quad g_2({\bf p}) = {\rm sin}\,
p_1, \quad g_3({\bf p}) = g_3^{(0)} + {\rm sin}\, p_3\nonumber\\&&
g_4({\bf p}) = \omega, \quad g_5({\bf p}) = m^{(0)} +
\sum_{a=1,2} (1 - {\rm cos}\, p_a), \quad b = const \label{ins2}
\end{eqnarray}
This system is able to describe qualitatively a certain class of topological insulators \cite{TIclassification}. Depending on the identification of matrices $\sigma^a$ entering matrices $\gamma^a$ of Eq. (\ref{G202}) with the operators of spin, isospin, or Bogolyubov spin, Eq. (\ref{ins2}) may describe different classes of topological insulators, including those with broken time reversal symmetry and/or broken particle - hole symmetry.
We assume, that the parameters entering the expression for the Green function satisfy $g^{(0)}_3 > 1$, $m^{(0)}> g^{(0)}_3-1$ while
$$\sqrt{(g^{(0)}_3 + 1)^2+(m^{(0)})^2} < b $$ and
$$\sqrt{(g^{(0)}_3 -1)^2+(m^{(0)} + 2)^2} > b$$

We will use Eqs. (\ref{Mpm}) and (\ref{G12d23}).
Let us denote
\begin{eqnarray}
{g}^\prime_1 = g_2, \quad {g}^\prime_2 = -g_1, \quad {g}^\prime_3 = g_4, \quad {g}^\prime_4 = \mp \Big(\sqrt{g_3^2({\bf p})+g_5^2({\bf p})} \pm b({\bf p})\Big)
\end{eqnarray}
and $\hat{g}^\prime_k = \frac{g^\prime_k}{g^\prime}$,  where $g^\prime = \sqrt{\sum_{k=1,2,3,4}[g^\prime_k]^2}$.
Then we have
$$\hat{g}^\prime_4({\bf p}) = \mp \frac{\sqrt{(g^{(0)}_3 + {\rm sin}\,p_3)^2+(m^{(0)} + \sum_{i=1,2}(1-{\rm cos}\,p_i))^2} \pm b}{\sqrt{(g^{(0)}_3 + {\rm sin}\,p_3)^2 + (m^{(0)} + \sum_{i=1,2}(1-{\rm cos}\,p_i))^2 + {\rm sin}^2\,p_1+ {\rm sin}^2\,p_2 + \omega^2}} $$
The above mentioned ranges of parameters guarantee, that
\begin{eqnarray}
\hat{g}^\prime_4({\bf p}) & = & 0, \quad {\bf p}\in \partial{\cal M} \quad (\omega \rightarrow \pm \infty)\nonumber\\
\hat{g}^\prime_4({\bf p}) & = & - 1, \quad \hat{g}^\prime_i({\bf p})  = 0\quad (k = 1,2,3),\quad  {\bf p} = (0,0,p_3,0),\quad p_3\in (-\pi,\pi)
\nonumber\\
\hat{g}^\prime_4({\bf p}) & = & \mp 1,\quad  \hat{g}^\prime_i({\bf p})  = 0\quad (k = 1,2,3),\quad  {\bf p} = (0,\pi,p_3,0),\quad p_3\in (-\pi,\pi) \nonumber\\
\hat{g}^\prime_4({\bf p}) & = & \mp 1,\quad  \hat{g}^\prime_i({\bf p})  = 0\quad (k = 1,2,3),\quad  {\bf p} = (\pi,0,p_3,0),\quad p_3\in (-\pi,\pi)\nonumber\\
\hat{g}^\prime_4({\bf p}) & = & \mp 1,\quad  \hat{g}^\prime_i({\bf p})  = 0\quad (k = 1,2,3),\quad  {\bf p} = (\pi,\pi,p_3,0),\quad p_3\in (-\pi,\pi)\label{listzerosI2}
\end{eqnarray}

We are able to use Eq. (\ref{Mpm}) if Eq. (\ref{cond00}) is fulfilled at $\omega \rightarrow \pm \infty$. This can be checked directly. We find, that again ${\cal M}^\prime_1 = {\cal M}^\prime_2 ={\cal M}^\prime_4 = 0$ while
\begin{eqnarray}
{\cal M}^\prime_3 &=&   \frac{2\pi}{2}- \frac{2\pi}{2} (-1)-\frac{2\pi}{2}(-1) - \frac{2\pi}{2} = 2 \pi
\end{eqnarray}
As well as for the considered above system with the $2\times 2$ Green function we deal with the AQHE current
\begin{equation}
 {j}^k_{Hall} = \cor{-} \frac{1}{2\pi\,a}\,\epsilon^{jk3}E_j
\end{equation}
(Here $a$ is the lattice spacing.)

As it was mentioned above, the nontrivial topology in the bulk of the topological insulators, which possess the AQHE, manifests itself in the appearance of the surface Fermi lines. Let us suppose, that the given insulator sample has the cubic form. Then the Fermi lines appear on the boundaries that belong to planes $(xz)$ and $(yz)$. Those Fermi lines are placed at ${\bf p} = (0,0,p_3,0),\quad p_3\in (-\pi,\pi)$. On the boundaries that belong to $(xy)$ plane there is the Fermi point at $p_1 = p_2 = 0$. This pattern may easily be derived using the Wigner transform of the Green function $\tilde{G}^{(0)}({\bf R},{\bf p})$, which depends on coordinate denending parameter $b({\bf R})$. It is assumed, that this parameter vanishes when we approach boundary.

\section{AQHE in the toy models of $3+1$ D  Weyl semimetals}
\label{SectWeylSem}

\subsection{Model W1}

Let us extend the above consideration to the models of Weyl semimetals.
First, let us start from the toy model of Weyl semimetal with the Green function ${\cal G}^{-1} = i \omega - H({\bf p})$ and the Hamiltonian of the form
\begin{equation}
H = {\rm sin}\,p_1\, \sigma^2 - {\rm sin}\, p_2 \, \sigma^1 - (m^{(0)}-{\rm cos}\,p_3 + \sum_{i=1,2}(1-{\rm cos}\,p_i)) \, \sigma^3
\end{equation}
This is the modification of the model I1 considered above.
For $m^{(0)}\in (0,1)$ the system contains the two Fermi points
\begin{equation}
{\bf K}_\pm = (0,0,\pm \beta,0), \quad \beta = {\rm arccos}\,m^{(0)}
\end{equation}
Although the Green function contains singularities, the integral in Eq. (\ref{nuGHall}) is convergent as we will see below. To be explicit, we first integrate over the momentum space with the small vicinities of the poles subtracted, and then consider the limit, when those vicinities are infinitely small.
We will be using the following expression for the Green function:
\begin{equation}
-i\sigma^3{\cal G}^{-1} = {\rm sin}\,p_1\, \sigma^1 + {\rm sin}\, p_2 \, \sigma^2 +  \omega \, \sigma^3 - i (m^{(0)} - {\rm cos}\,p_3+ \sum_{i=1,2}(1-{\rm cos}\,p_i))
\end{equation}
We have
$$\hat{g}_4({\bf p}) = \frac{(m^{(0)} - {\rm cos}\,p_3 + \sum_{i=1,2}(1-{\rm cos}\,p_i))}{\sqrt{(m^{(0)}- {\rm cos}\,p_3 + \sum_{i=1,2}(1-{\rm cos}\,p_i))^2+ {\rm sin}^2\,p_1+ {\rm sin}^2\,p_2 + \omega^2}} $$
and
\begin{eqnarray}
\hat{g}_4({\bf p}) & = & 0, \quad {\bf p}\in \partial{\cal M}\nonumber\\
\hat{g}_4({\bf p}) & = & -1, \quad \hat{g}_i({\bf p})  = 0\quad (k = 1,2,3),\quad  {\bf p} = (0,0,p_3,0),\quad p_3\in (-\beta,\beta)
\nonumber\\
\hat{g}_4({\bf p}) & = & 1, \quad \hat{g}_i({\bf p})  = 0\quad (k = 1,2,3),\quad  {\bf p} = (0,0,p_3,0),\quad p_3\in (-\pi,-\beta)\cup (\beta,\pi)\nonumber\\
\hat{g}_4({\bf p}) & = & 1,\quad  \hat{g}_i({\bf p})  = 0\quad (k = 1,2,3),\quad  {\bf p} = (0,\pi,p_3,0),\quad p_3\in (-\pi,-\pi) \nonumber\\
\hat{g}_4({\bf p}) & = & 1,\quad  \hat{g}_i({\bf p})  = 0\quad (k = 1,2,3),\quad  {\bf p} = (\pi,0,p_3,0),\quad p_3\in (-\pi,-\pi)\nonumber\\
\hat{g}_4({\bf p}) & = & 1,\quad  \hat{g}_i({\bf p})  = 0\quad (k = 1,2,3),\quad  {\bf p} = (\pi,\pi,p_3,0),\quad p_3\in (-\pi,-\pi)\label{listzerosW1}
\end{eqnarray}
Therefore ${\cal M}^\prime_1 = {\cal M}^\prime_2 ={\cal M}^\prime_4 = 0$ while
\begin{eqnarray}
{\cal M}^\prime_3 &=&  -\frac{2\pi-2\beta}{2} + \frac{2\beta}{2}- \frac{2\pi}{2} (-1)-\frac{2\pi}{2}(-1) - \frac{2\pi}{2} = 2\beta
\end{eqnarray}
Thus we come to the expression for the AQHE current
\begin{equation}
{j}^k_{Hall} =\cor{-} \frac{\beta}{2\pi^2}\,\epsilon^{jk3}E_j,\label{HALLj3dp}
\end{equation}
From the above expression it is obvious, that the contributions of all zeros of ${g}_k$ ($k=1,2,3$) listed in Eq. (\ref{listzerosW1}) are important. Nevertheless, our result coincides with the one of the naive low energy effective field theory \cite{Zyuzin:2012tv}. The coefficient in Eq. (\ref{HALLj3dp}) has also been calculated using the technique of Berry curvature in \cite{AQHEWeyl}.

It is also worth mentioning, that in the particular case of the Weyl semimetal, when the Green function contains poles Eq. (\ref{nuGHall}) is not the topological invariant. It still gives the AQHE current, though. But we should remember, that turning on interactions we change the value of ${\cal M}^\prime_k$, thus leading to the continuous renormalization of the value $\beta$ that marks the position of the Fermi point. Notice, that even in the presence of a pole Eq. (\ref{nuGHall}) remains invariant under those continuous changes of the Green function, which do not alter the positions of the Fermi points and for which
$$\int {\rm Tr} \left(
\{[\delta {\cal G}^{-1}] {\cal G}] \}
  d  {\cal G}^{-1}\wedge
  d  {\cal G}\right) = 0$$
where the integral is taken along the boundary of the small vicinity of the pole.
This follows immediately from the derivation of Appendix B.

It is worth mentioning, that the first order in the derivative expansion does not allow to calculate the derivative of electric current. That's why we cannot simply differentiate Eq. (\ref{HALLj3dp}) in order to calculate such a derivative. Both dependence of $\beta$ on $\bf R$ and the second order in the derivative expansion should be applied  to calculate $\partial j$ (that should vanish in equilibrium due to the gauge invariance).

\subsection{Model W2}

Let us consider the more complicated model of Weyl semimetal, which represents the deformation of the insulator model I2 with the Green function of the form of Eq. (\ref{G202}). We imply, that the functions $g_a$ and $b$ entering Eq. (\ref{G202}) have the form of Eq. (\ref{ins2}). But we assume different relations between the parameters of the model. Now we require, that the parameters entering the expression for the Green function satisfy $g^{(0)}_3 > \sqrt{b^2-(m^{(0)})^2}>g^{(0)}_3-1>0$ while
$$\sqrt{(g^{(0)}_3 + {\rm sin}\, \beta_\pm)^2+(m^{(0)})^2} = b $$
$$\sqrt{(g^{(0)}_3 + {\rm sin}\, p_3)^2+(m^{(0)})^2} < b, \quad p_3\in (\beta_-,\beta_+), $$
$$\sqrt{(g^{(0)}_3 + {\rm sin}\, p_3)^2+(m^{(0)})^2} > b, \quad p_3\in (-\pi,\beta_-)\cup (\beta_+,\pi) $$
The given model contains the two Fermi points
\begin{equation}
{\bf K}_\pm = (0,0, \beta_\pm,0)
\end{equation}

Instead of Eq. (\ref{listzerosI2}) we have
\begin{eqnarray}
\hat{g}^\prime_4({\bf p}) & = & 0, \quad {\bf p}\in \partial{\cal M} \quad (\omega \rightarrow \pm \infty)\nonumber\\
\hat{g}^\prime_4({\bf p}) & = & - 1, \quad \hat{g}^\prime_i({\bf p})  = 0\quad (k = 1,2,3),\quad  {\bf p} = (0,0,p_3,0),\quad p_3\in (\beta_-,\beta_+)
\nonumber\\
\hat{g}^\prime_4({\bf p}) & = & \mp 1, \quad \hat{g}^\prime_i({\bf p})  = 0\quad (k = 1,2,3),\quad  {\bf p} = (0,0,p_3,0),\quad p_3\in (-\pi,\beta_-)\cup (\beta_+,\pi)\nonumber\\
\hat{g}^\prime_4({\bf p}) & = & \mp 1,\quad  \hat{g}^\prime_i({\bf p})  = 0\quad (k = 1,2,3),\quad  {\bf p} = (0,\pi,p_3,0),\quad p_3\in (-\pi,\pi) \nonumber\\
\hat{g}^\prime_4({\bf p}) & = & \mp 1,\quad  \hat{g}^\prime_i({\bf p})  = 0\quad (k = 1,2,3),\quad  {\bf p} = (\pi,0,p_3,0),\quad p_3\in (-\pi,\pi)\nonumber\\
\hat{g}^\prime_4({\bf p}) & = & \mp 1,\quad  \hat{g}^\prime_i({\bf p})  = 0\quad (k = 1,2,3),\quad  {\bf p} = (\pi,\pi,p_3,0),\quad p_3\in (-\pi,\pi)\label{listzerosW2}
\end{eqnarray}

In Eq. (\ref{nuGHall}) we represent ${\cal M}_l$ as an integral over the component $p_l$ of the topological invariant of the $2+1$ D system, which depends on $p_l$ as on parameter. In the presence of the pole the integral is ill - defined at the positions of poles and requires regularization. We consider the integration over $p_l$ with the small vicinities of the $l$ - th coordinates of the poles subtracted. For each value of $p_l$, which does not belong to this interval the integral over the other components of $\bf p$ is well - defined. Therefore, we are able to use Eq. (\ref{Mpm}) if Eq. (\ref{cond00}) is fulfilled at $\omega \rightarrow \pm \infty$ (which may be checked easily).
The values of ${\cal M}^\prime_{j,\pm}$ are given by the integrals in Eq. (\ref{calMproj}) over $dp_j$, where as it was explained above, the small vicinities $(p^{(a)}_j-\epsilon,p^{(a)}_j+\epsilon)$ of the $j$ - th coordinates of the poles ${\bf p}^{(a)}$, $a=1,2$ are subtracted. At the end of the calculations we take the limit $\epsilon \rightarrow 0$.

This way we find, that ${\cal M}^\prime_1 = {\cal M}^\prime_2 ={\cal M}^\prime_4 = 0$ while
\begin{eqnarray}
{\cal M}^\prime_3 &=&  -\frac{2\pi-(\beta_+ -\beta_-)}{2} + \frac{(\beta_+ -\beta_-)}{2}- \frac{2\pi}{2} (-1)-\frac{2\pi}{2}(-1) - \frac{2\pi}{2} = (\beta_+ -\beta_-)
\end{eqnarray}
The AQHE current reads
\begin{equation}
{j}^k_{Hall} = \frac{\cor{\beta_--\beta_+}}{4\pi^2}\,\epsilon^{jk3}E_j,\label{HALLj3dp}
\end{equation}

In order to illustrate how the bulk - boundary correspondence works here, let us consider what happens to the Fermi points when we aproach the boundary of the sample. This approaching is described by the Wigner transform of the Green function $\tilde{G}^{(0)}({\bf R},{\bf p})$, which depends on the coordinates ${\bf R}$ through the dependence of parameter $b$ on ${\bf R}$. We assume, that out of the bulk of the semimetal $b({\bf R}) = 0$, and this parameter is increased inside the semimetal.   In this situation the derivative expansion in the presence of external electric field becomes more complicated near boundary than inside of the bulk. However, we do not discuss here the corresponding expression for the electric current. As for the Fermi points, one can easily see, that the two Fermi points ${\bf K}_\pm$ approach each other, and finally annihilate while we are aproaching boundary. Since this process occurs in the small vicinity of the boundary, it results in the appearance of surface states of zero energy that connect the two bulk Fermi points. These surface states are the Fermi arcs.


\section{The absence of bulk chiral magnetic current}
\label{SectCME}

The conventional expression for the CME current reads
$j^{k}_{CME} = \frac{\mu_5}{4\pi^2}\,\epsilon^{ijk4}\, A_{ij}$, where $\mu_5$ is the chiral chemical potential.
 One may expect, that such an expression will follow from Eq. (\ref{calM}): we need to substitute ${\bf A}=0$ into Eq. (\ref{nuG}) in the linear response approximation. Then one might expect that ${\cal M}_4 = \mu_5$ for the system of a single massless Dirac fermion and a similar expression may take place for the other systems with gapless fermions (or, nearly gapless fermions).

Let us consider the situation, when vector gauge field $A_k({\bf R})$ has the nonzero components with $k=1,2,3$ that do not depend on imaginary time. There may exist many different definitions of $\mu_5$. The most straightforward way is to consider the following expression for the fermion Green function:
 \begin{equation}
 {\cal G}^{}({\bf p}) = \Big(\sum_{k}\gamma^{k} g_{k}({\bf p}) + i\gamma^4 \gamma^5 \mu_5 - i m({\bf p})\Big)^{-1}\label{G2}
 \end{equation}
In the limit of vanishing chiral chemical potential it is reduced to the form of Eq. (\ref{G10}),
 where $\gamma^k$ are Euclidean Dirac matrices while $g_k({\bf p})$ and $m({\bf p})$ are the real - valued functions, $k = 1,2,3,4$.
We define $\gamma^5$  in chiral representation as ${\rm diag}(1,1,-1,-1)$. We may substitute ${\cal G}$ of Eq. (\ref{G2}) into Eq. (\ref{calM}) instead of $\tilde{G}({\bf R},{\bf p})$ while dealing with the linear response to the external magnetic field.

We will apply our methodology to the investigation of the theory with compact Brillouin zone. We imply, that momentum space can be represented as $R^1 \otimes \Omega$ or $S^1\otimes \Omega$, where $\Omega$ is the compact $3D$ Brillouin zone. First, we assume, that the Green functions do not have zeros or poles, which means, that the fermions are gapped (massive). However, at the end of the calculations the limit of vanishing mass may always be considered. We need, though, that the inclusion of the chiral chemical potential does not produce poles in the Green function. For the Green function of the form of Eq. (\ref{G2}) this may be proved as follows. The poles of the Green function appear as the solutions of the following equation
\begin{equation}
g^2_4({\bf p}) + \Big(\mu_5 \pm \sqrt{g_1^2({\bf p})+g_2^2({\bf p})+g_3^2({\bf p})}\Big)^2 + m^2({\bf p}) = 0
\end{equation}
One can easily see, that if functions $m({\bf p})$ and $g_4({\bf p})$ never vanish simultaneously, then the chiral chemical potential of Eq. (\ref{G2}) cannot lead to the appearance of the poles\footnote{This is in contrast to the case of the ordinary chemical potential, in which the poles of the Green function may appear if chemical potential exceeds the gap.} of $\cal G$. To illustrate this general pattern let us consider the Green function for the toy model of topological insulator inspired by the lattice regularization with free Wilson fermions:
\begin{equation}
g_4 = p_4\equiv \omega, \quad g_k({\bf p}) = {\rm sin}\,
p_k\, (k=1,2,3),  \quad m({\bf p}) = m^{(0)} +
\sum_{a=1,2,3} (1 - {\rm cos}\, p_a)\label{Wilsonlike}
\end{equation}
One can see, that for the positive values of bare mass parameter $m^{(0)}$ the values of $m({\bf p})$ do not vanish.
${\cal M}_4$ is topological invariant, i.e. it is robust to any variations of the Green function as long as the singularities are not encountered (for the proof see Appendix B). The introduction of chiral chemical potential is the particular case of such a variation. Actually, for the Green function of the form of Eq. (\ref{G10}) ${\cal M}_4 = 0$, which may be proved easily using the properties of gamma matrices. Therefore, if the nonzero value of $\mu_5$ does not cause the appearance of the poles of the Green function, then ${\cal M}_4 = 0$ for $\mu_5 \ne 0$.

The mentioned above consideration is sufficient to prove, that there is no CME in the lattice regularized relativistic quantum field theory because the Wilson fermions with $m^{(0)}>0$ represent the typical lattice regularization. However, in the solid state systems the situation may be different from the one considered above. This may be illustrated by the model of Eq. (\ref{Wilsonlike}) for $m^{(0)} \in (-2,0)$. The corresponding value of the topological invariant $\tilde{\cal N}_5$ introduced in \cite{VZ2012} is nonzero. In this case the nonzero value of $\mu_5$ causes the appearance of the Fermi lines. Those Fermi lines appear as the solution of the system of equations
\begin{eqnarray}
3 + m^{(0)} & = & {\rm cos}\, p_1 + {\rm cos}\, p_2 + {\rm cos}\, p_3  \nonumber \\
|\mu_5| & = & \sqrt{ {\rm sin}^2\, p_1 + {\rm sin}^2\, p_2 + {\rm sin}^2\, p_3  }\label{mu50}
\end{eqnarray}
This system has the nontrivial solution at the sufficiently large value of $\mu_5$.
This manyfold of the zeros of the Green function is marginal because it has the dimension smaller, than that of the ordinary Fermi surface caused by the finite values of the ordinary chemical potential.

The value of ${\cal M}_4$ is given by
\begin{eqnarray}
 {\cal M}_{4} &=& - \frac{i}{2}\int d \omega \tilde{\cal N}_3(\omega),\label{calM40}\\
 \tilde{\cal N}_3(\omega) & = &  \frac{1}{24 \pi^2}\epsilon_{ijk4} {\rm Tr}\,  \int_{\Omega} d^3 p \Big( {\cal G} \partial^i {\cal G}^{-1} \Big)\nonumber\\&&\Big( {\cal G}\partial^j {\cal G}^{-1} \Big)\Big({\cal G} \partial^k {\cal G}^{-1}\Big) \label{F4B0}
\end{eqnarray}
For $\omega \ne 0$ the quantity $\tilde{\cal N}_3(\omega)$ is the topological invariant. We may perform the deformation, which brings $\mu_5$ to zero. For $\mu_5=0$ we may easily prove, that $\tilde{\cal N}_3(\omega) = 0$. Therefore, at $\omega \ne 0$ the value of $\tilde{\cal N}_3(\omega) $ vanishes at nonzero values of $\mu_5$. The integral over $\omega$ in Eq. (\ref{calM40}) is to be regularized as
\begin{equation}
\int = {\rm lim}_{\epsilon \rightarrow 0}\Big(\int_{-\infty}^{-\epsilon} + \int_\epsilon^{\infty}\Big)\label{epsilon}
 \end{equation}
 The limit $\epsilon \rightarrow 0$ gives ${\cal M}_4 = 0$. The same conclusion may be drawn for the finite temperature $T$, when instead of the integral over $\omega$ the sum over the Matsubara frequencies $\omega_n = T \pi (2n+1)\ne 0$ appears. At each value of $\omega_n$  the expression of Eq. (\ref{F4B0}) is topological invariant, and therefore it is equal to zero.

One can easily find that the considered above pattern appears in the other noninteracting models of topological insulators based on Eq. (\ref{G2}). If topology is nontrivial, then the nonzero value of chiral chemical potential causes the appearance of the Fermi lines. The integral in Eq. (\ref{calM40}) is to be regularized through Eq. (\ref{epsilon}) and gives vanishing value of ${\cal M}_4$. This proves the absence of the equilibrium CME in such systems.

The above consideration referred to gapped fermions. The Dirac semimetals may be considered as the limiting case of vanishing (or, better to say, of very small) value of the gap. This limit does not change the above conclusion about the absence of the CME. Moreover, following \cite{HAL} we may consider the system of gapless fermions from the very beginning. In the non - interacting system the poles of the Green function appear at $\omega = 0 $ only. This requires the same regularization of Eq. (\ref{epsilon}) for the expression of ${\cal M}_4$ and gives rise to the same conclusion on the absence of CME.

\section{Conclusions and discussion}
\label{SectConcl}

In the present paper we demonstrated how momentum space topology may be applied to the analysis of the anomalous transport related to the non - dissipative electric current. We concentrate on the linear response of electric current to external electric and magnetic fields. The response to external electric field leads to the appearance of the quantum Hall effect while the response to external magnetic field was associated in certain publications with the equilibrium chiral magnetic effect.

We show that the corresponding currents are proportional to the momentum space topological invariants. Our methodology is based on the derivative expansion applied to the Wigner transform of the two - point Green functions. We introduce the slowly varying external gauge field directly to the momentum space formulation of the lattice models of solid state physics and of the lattice regularization of continuous QFT. In this representation the external gauge field appears as a pseudo - differential operator ${\bf A}(i\partial_{\bf p})$. This way of the incorporation of the external field to the theory is not useful for its numerical simulations, but is convenient for the analytical derivations.
As it was mentioned above, the response of electric current to external field strength is expressed through the topological invariant in momentum space. This relation allows to describe the anomalous quantum Hall effect both in the $2+1$ D and in the $3+1$ D systems. First of all we reproduce the conventional expression for the $2+1$ D Hall conductivity through the topological invariant $\tilde{\cal N}_3$ (in the classification of \cite{Volovik2003}).  It is demonstrated, how this invariant may be calculated in practise in the systems with the $2\times 2$ Green function. Besides, we reproduce the conventional expression of the $2+1$ D Hall conductivity through the integral over Berry curvature \cite{TTKN,Hall3DTI}.

Next, we extend our consideration to the $3+1$ D models. We describe the way to calculate the corresponding topological invariants for the wide class of systems with $2\times 2$ and $4\times 4$ Green functions (including those with the broken time reversal symmetry). This method is illustrated by the consideration of the two particular toy models of topological insulators. We demonstrate, that the resulting Hall current is proportional to one of the vectors of reciprocal lattice (which was also proposed in  \cite{Klinkhamer:2004hg}). We also show, how the bulk boundary correspondence may be ivestigated using the Wigner transform of the Green functions. It is pointed out, that the insulator state, which admits AQHE is accompanied by the appearance of the topologically protected surface Fermi lines incident at the boundary of the insulator.

Further, we consider the $3+1$ D Weyl semimetals. The two toy models of Weyl semimetals with the $2\times 2$ and $4\times 4$ Green functions are investigated in details. The conventional expressions for AQHE currents obtained earlier using the effective low energy field theory \cite{semimetal_effects11} are reproduced in both cases. Our findings here are in line with the recent results on AQHE reported in \cite{nogo,AQHEWeyl}. We also briefly discuss how the surface Fermi arcs connecting the bulk Weyl points manifest themselves on the language of the Wigner transformation of the Green functions.

Our analysis of the equilibrium chiral magnetic effect demonstrates that the corresponding topological invariant entering the expression for the CME current does not depend on the value of chiral chemical potential for the systems with compact Brillouin zone and without poles or zeros of the Green function. Therefore, we conclude, that the gapped  solid state systems do not possess the equilibrium bulk chiral magnetic effect. The same conclusion refers to the semimetals, which may be considered as the insulator with very small gap. Such a limit does not change the independence of the mentioned above topological invariant on chiral chemical potential even when the value of $\mu_5$ exceeds the gap. Thus, we confirm the recent numerical results \cite{Valgushev:2015pjn,Buividovich:2015ara,Buividovich:2015ara,Buividovich:2014dha,Buividovich:2013hza}, which led to the same conclusion.

To conclude, in all considered cases the proposed methodology reproduces the known results on the AQHE and CME currents. However, the class of models analysed here is more wide. Unlike the more popular method, which utilizes Berry curvature, we use the topological invariants composed of the Green functions \cite{Volovik2003}. This allows us to investigate the topological contribution to electric current in the interacting systems, when the Green function does not have the simple form ${\cal G}^{-1} = i\omega - \hat{H}$ with the Hamiltonian $\hat{H}$ and imaginary frequency $\omega$. We relate both AQHE and CME currents to such invariants composed of the Wigner transform of the Green functions. This representation is unique and allows to describe both bulk AQHE and the bulk - boundary correspondence. The main advantage of the proposed methodology is the possibility to describe systematically the anomalous transport in the wide class of models, including those with the interactions turned on. The proposed method of the calculation of the mentioned topological invariants may, in principle, be extended to the more complicated systems.

The author kindly acknowledges useful discussions with G.E.Volovik and M.N.Chernodub.

\section*{Appendix A. Wigner transformation of the Green function}
\label{SectWignerl}

Here we repeat the description of the Wigner transformation in momentum space proposed in \cite{HAL}. \cor{Notice that the expressions given below are valid for the systems with sufficiently weak inhomogeneity that may be neglected at the distance of the lattice spacing.} Let us consider the $d+1 = D$ dimensional lattice model of solid state physics (or the lattice regularization of the continuum QFT) with the Green function ${\cal G}({\bf p}_1,{\bf p}_2)$, which obeys equation
\begin{equation}
\hat{\cal Q}(i \partial_{{\bf p}_1},{\bf p}_1)G({\bf p}_1,{\bf p}_2) = |{\cal M}| \delta^{(D)}({\bf p}_1-{\bf p}_2)\label{QGlA}
\end{equation}
Here  $\hat{\cal Q}$ is Hermitian operator - valued function. Wigner decomposition in momentum space takes the form
\begin{equation}
 \tilde{G}({\bf R},{\bf p}) = \int \frac{d^D{\bf P}}{|{\cal M}|} e^{i {\bf P} {\bf R}} G({\bf p}+{\bf P}/2,{\bf p}-{\bf P}/2)\label{WlA}
\end{equation}
Below we will prove that the Wigner transform of the Green function obeys Groenewold equation
\begin{equation}
 {\cal Q}({\bf R},{\bf p})e^{\frac{i}{2}(\overleftarrow{\partial}_{\bf R}\overrightarrow{\partial}_{\bf p} - \overleftarrow{\partial}_{\bf p}\overrightarrow{\partial}_{\bf R})}\tilde G({\bf R},{\bf p})  = 1 \label{idBl}
\end{equation}
Here $\cal Q$ is the Weyl symbol of operator $\hat{\cal Q}$, that is the function of real numbers rather than of the operators.
In \cite{Weyl,berezin} it is defined as
\begin{equation}
 {\cal Q}({\bf R},{\bf p}) = \int {d^D {\bf K}} {d^D{\bf P}} e^{i {\bf P} {\bf R}} \delta({\bf p}-{\bf P}/2 - {\bf K}) \hat{\cal Q}(i {\partial}_{\bf K},{\bf K})   \delta({\bf p}+{\bf P}/2 - {\bf K})\label{W000}
\end{equation}
The derivation of Eq. (\ref{idBl}) was given, for example, in \cite{berezin}.

The further consideration will be based on the different definition of Weyl symbol. Its equivalence to that of Eq. (\ref{W000}) follows from the fact that it obeys the same equation Eq. (\ref{idBl}).
To be explicit, we determine relation between the function ${\cal Q}({\bf r},{\bf p})$ (of real - valued vectors ${\bf r}$ and ${\bf p}$) and the function $\hat{\cal Q}(\hat{\bf r}, {\bf p})$ (of the operators ${\bf p}$ and $\hat{\bf r} = i \partial_{\bf p}$) through equation
\begin{eqnarray}
&& \int d^D {\bf X} d^D {\bf Y}\, f({\bf X},{\bf Y})\, {\cal Q}(-i \overleftarrow{\partial}_{\bf Y}+i\overrightarrow{\partial}_{\bf X},{\bf X}/2+{\bf Y}/{2})\, h({\bf X},{\bf Y})\nonumber\\ && =  \int  d^D {\bf X}d^D {\bf Y}\, f({\bf X},{\bf Y}) \hat{\cal Q}\Big(i \partial_{\bf X}+ i\partial_{\bf Y},{\bf X}/2+ {\bf Y}/{2}\Big) \, h({\bf X},{\bf Y}) \label{corrl}
\end{eqnarray}
which is valid for arbitrary functions $f({\bf X},{\bf Y})$ and $h({\bf X},{\bf Y})$ defined on momentum space ${\bf X},{\bf Y}\in {\cal M}$. The derivatives  $\overrightarrow{\partial}_{\bf X}$ and $\overleftarrow{\partial}_{\bf Y}$ inside the arguments of $\cal Q$ act only outside of this function, i.e. $\overleftarrow{\partial}_{\bf Y}$ acts on $f({\bf X},{\bf Y})$ while $\overrightarrow{\partial}_{\bf X}$  acts on $h({\bf X},{\bf Y})$. At the same time
the derivatives without arrows act as usual operators, i.e. not only right to the function $\hat{\cal Q}$, but inside it as well. Notice, that $\frac{\partial}{\partial ({\bf X}/2+ {\bf Y}/{2})} = \partial_{\bf Y} + {\partial_{\bf X}}$ and $\frac{\partial}{\partial ({\bf X}/2 - {\bf Y}/{2})} = \partial_{\bf X} - {\partial_{\bf Y}}$. We may rewrite Eq. (\ref{corrl}) as
\begin{eqnarray}
&& \int d^D {\bf X} d^D {\bf Y}\, f({\bf X},{\bf Y})\, {\cal Q}(-i \overleftarrow{\partial}_{\bf Y}+i\overrightarrow{\partial}_{\bf X},{\bf X}/2+{\bf Y}/{2})\, h({\bf X},{\bf Y})\nonumber\\ && = -2 \int d^D {\bf Q} d^D {\bf K}\, f({\bf Q}+{\bf K},{\bf Q}-{\bf K}) \hat{\cal Q}\Big(i \partial_{\bf Q},{\bf Q}\Big) \, h({\bf Q}+{\bf K},{\bf Q}-{\bf K}) \label{corrl2}
\end{eqnarray}

The given correspondence between operator $\hat{\cal Q}$ and its symbol $\cal Q$ takes the simple form in certain particular cases.
For example, if $\hat{\cal Q} = ({\bf p} - {\bf A}(\hat{\bf r}))^2 = {\bf p}^2 + {\bf A}^2(\hat{\bf r}) + i \Big(\partial^k A_k(\hat{\bf r})\Big) - 2 {\bf A}(\hat{\bf r}){\bf p}$ (recall, that  $\hat{\bf r}$ is operator equal to $i\partial_{\bf p}$), then  ${\cal Q} = {\bf p}^2 + {\bf A}^2({\bf r}) - 2 {\bf A}({\bf r}){\bf p}$. Besides, if $\hat{\cal Q}$ has the form
\begin{equation}
\hat{\cal Q}(\hat{\bf r},{\bf p}) = {\cal F}({\bf p} - {\bf A}(\hat{\bf r}))
\end{equation}
then
\begin{equation}
{\cal Q}({\bf r},{\bf p}) = {\cal F}({\bf p} - {\bf A}({\bf r})) + O([\partial_i A_j]^2)\label{QF}
\end{equation}
Here $O([\partial_i A_j]^2)$ may contain the terms with the second power of the derivatives of $\bf A$  and the terms higher order in derivatives.

In order to prove Eq. (\ref{QF}) let us represent the function ${\cal F}({\bf p} - {\bf A}(\hat{\bf r})) = \sum_{n}{\cal F}_{i_1...i_n}(p_{i_1} - A_{i_1}(i\partial_{\bf p}))...(p_{i_n} - A_{i_n}(i\partial_{\bf p}))$ as a series in powers of its arguments (${\cal F}_{i_1...i_n}$ are Hermitian operators that do not depend on $\bf p$).  Operator $\hat{Q}$ is Hermitian, therefore, the operator in the first row of Eq. (\ref{corrl2}) should also be Hermitian.

Suppose, that function $\cal Q$ is expanded in powers of ${\bf Q}=({\bf X}+{\bf Y})/2$ and $-i \overleftarrow{\partial}_{\bf Y}+i\overrightarrow{\partial}_{\bf X}$ as follows
\begin{equation}
{\cal Q}(-i \overleftarrow{\partial}_{\bf Y}+i\overrightarrow{\partial}_{\bf X},{\bf X}/2+{\bf Y}/{2}) = \sum q_{i_1...i_n;j_1...j_m;k_1...k_l}(-i \overleftarrow{\partial}_{Y_{i_1}})...(-i \overleftarrow{\partial}_{Y_{i_n}}) Q_{j_1}...Q_{j_m}(i\overrightarrow{\partial}_{X_{k_1}})(i\overrightarrow{\partial}_{X_{k_l}})\label{QQ}
\end{equation}
In this expression inside the first row of Eq. (\ref{corrl}) we may be substitute $-i\overleftarrow{\partial}_{Y_{i}}$ by  $i{\partial}_{Y_{i}}$ and $i\overrightarrow{\partial}_{X_{k}}$ by $i{\partial}_{X_{k}}$. Because the second row in Eq. (\ref{corrl}) is symmetric under the interchange of $\bf X$ and $\bf Y$, we have $q_{i_1...i_n;j_1...j_m;k_1...k_l}=q_{k_1...k_l;j_1...j_m;i_1...i_n}$. For the same reason Eq. (\ref{QQ}) is invariant under the interchange ${\bf X}\leftrightarrow {\bf Y}$. Then the change of $q_{...}$ by its Hermitian conjugate $q^+_{...}$ is equivalent to the Hermitian conjugation of the whole expression. This demonstrates that coefficients $q_{...}$ are Hermitian.

Let us suppose, that ${\cal Q}$ is linear in the derivative of ${\bf A}$. The linear term appears as a product of a certain combination of ${\cal F}_{...}$ and the commutator $[{p}_k,{\bf A}(i\partial_{\bf p})] = - i (\partial_k {\bf A})$. Therefore, it would lead to the appearance of imaginary unity in the expression for $q_{...}$ as a combination of ${\cal F}_{...}$, which means that $q_{...}$ is not Hermitian. Therefore, we come to the contradiction, which proves the non - appearance of the terms linear in the derivatives of ${\bf A}$ in the expression for ${\cal Q}({\bf r},{\bf p})$.

Now let us prove Eq. (\ref{idBl}). In order to do this let us substitute Eq. (\ref{WlA}) into Eq. (\ref{idBl}).
Argument of the exponent in Eq. (\ref{idBl}) acts on $\cal Q$ as follows:
\begin{eqnarray}
1&=&\int \frac{d^D{\bf P}}{|{\cal M}|}  {\cal Q}({\bf R}+\frac{i}{2}\overrightarrow{\partial}_{\bf p},{\bf p}-\frac{i}{2}\overrightarrow{\partial}_{\bf R}) \nonumber\\&& e^{i {\bf P} {\bf R}} G({\bf p}+{\bf P}/2,{\bf p}-{\bf P}/2)
\end{eqnarray}
In this expression the derivatives  $\overrightarrow{\partial}_{\bf p}$ and $\overrightarrow{\partial}_{\bf R}$ inside the arguments of $\cal Q$ act only outside of this function, i.e. on $e^{i {\bf P} {\bf R}} G({\bf p}+{\bf P}/2,{\bf p}-{\bf P}/2)$ and do not act inside the function $\cal Q$, i.e. on $\bf p$ and $\bf R$ in its arguments.
This gives
\begin{eqnarray}
1&=&\int \frac{d^D{\bf P}}{|{\cal M}|} e^{i {\bf P} {\bf R}} {\cal Q}(-i \overleftarrow{\partial}_{\bf P}+\frac{i}{2}\overrightarrow{\partial}_{\bf p},{\bf p}+\frac{\bf P}{2})\\  && G({\bf p}+{\bf P}/2,{\bf p}-{\bf P}/2)  \nonumber
\end{eqnarray}
Integrating by parts we arrive at
\begin{equation}
\int \frac{d^D{\bf P}}{|{\cal M}|} e^{i {\bf P} {\bf R}} \hat{\cal Q}(i {\partial}_{\bf P}+\frac{i}{2}{\partial}_{\bf p},{\bf p}+\frac{\bf P}{2})   G({\bf p}+{\bf P}/2,{\bf p}-{\bf P}/2)  = 1\nonumber
\end{equation}
(Eq. (\ref{corrl} is used.) Next, we take into account Eq. (\ref{corrl2}) and apply the inverse Wigner transform, which leads us to  Eq. (\ref{QGlA}).

\section*{Appendix B. Topological invariant responsible for the linear response of electric current to external  field strength}

Let us consider the following expression for the coefficient entering the linear response of electric current to external magnetic field:
\begin{eqnarray}
 {\cal M}_{l} &=& - \frac{i}{48 \pi^2}\epsilon_{ijkl} {\rm Tr}\,  \int_{\cal M} d^4 p \Big( {\cal G} \partial^i {\cal G}^{-1} \Big)\Big( {\cal G}\partial^j {\cal G}^{-1} \Big)\Big({\cal G} \partial^k {\cal G}^{-1}\Big) \label{F3B}
\end{eqnarray}
If $\cal M$ has the form of the product $R^1\otimes \Omega$, where $\Omega$ is the compact 3D Brillouin zone, while $R^1$ is the imaginary frequency line (or if ${\cal M} = S^1\otimes \Omega$, which takes place if the time evolution is discretized), then for $l=4$ we may rewrite this quantity as follows:
\begin{eqnarray}
 {\cal M}_{4} &=& - \frac{i}{2}\int dp^4 \tilde{\cal N}_3(p^4),\\
 \tilde{\cal N}_3(p^4) & = &  \frac{1}{24 \pi^2}\epsilon_{ijk4} {\rm Tr}\,  \int_{\Omega} d^3 p \Big( {\cal G} \partial^i {\cal G}^{-1} \Big)\Big( {\cal G}\partial^j {\cal G}^{-1} \Big)\Big({\cal G} \partial^k {\cal G}^{-1}\Big) \label{F4B}
\end{eqnarray}
Here for the fixed value of $p^4$ we encounter the expression for the topological invariant in the 3D Brillouin zone. Green function $\cal G$ should be considered here as the function of the 3 arguments $p^1,p^2,p^3$ while $p^4$ is to be considered as a parameter.

Notice, that for the Green function of the form of Eq. (\ref{G10}) the value of $\tilde{\cal N}_3(p^4)$ is equal to zero.  One might naively think, that the deviation of the Green function from the form of Eq. (\ref{G10}) - say, of the form of Eq. (\ref{G2}) may change the expressions for $\tilde{\cal N}_3$ and ${\cal M}_4$. Below we will demonstrate, that this does not occur as long as we deal with the compact Brillouin zone and regular Green functions.

Let us consider arbitrary variation of the Green function:
${\cal G} \rightarrow {\cal G} + \delta {\cal G}$. Then expression
for $\tilde{\cal N}_3$ is changed as follows:
\begin{eqnarray}
  \delta \tilde{\cal N}_3  & = &
 - \frac{3}{24  \pi^2} \int_{} {\rm Tr} \left( \{[\delta {\cal G}]
  d  {\cal G}^{-1}+{\cal G}
  d  [\delta {\cal G}^{-1}]\}\wedge {\cal G}
  d  {\cal G}^{-1}\wedge {\cal G}
  d  {\cal G}^{-1}\right)\nonumber\\&=&
-\frac{3}{24  \pi^2} \int_{} {\rm Tr} \left( \{-{\cal
G} [\delta {\cal G}^{-1}]{\cal G}
  d  {\cal G}^{-1}+{\cal G}
  d  [\delta {\cal G}^{-1}]\}\wedge {\cal G}
  d  {\cal G}^{-1}\wedge {\cal G}
  d  {\cal G}^{-1}\right) \nonumber\\&=&
\frac{3}{24  \pi^2} \int_{} {\rm Tr} \left(
\{[\delta {\cal G}^{-1}] [d{\cal G}] +
  d  [\delta {\cal G}^{-1}]{\cal G}\}\wedge
  d  {\cal G}^{-1}\wedge
  d  {\cal G}\right) \nonumber\\&=&
\frac{3}{24  \pi^2} \int_{} d \, {\rm Tr} \left(
\{[\delta {\cal G}^{-1}] {\cal G}] \}
  d  {\cal G}^{-1}\wedge
  d  {\cal G}\right) = 0 \label{dNP}
\end{eqnarray}
That's why  we proved that $\tilde{\cal N}_3$ and ${\cal M}_4$ are the topological invariants.

In the similar way it may be proved, that ${\cal M}_l$ is the topological invariant for $l\ne 4$ if ${\cal M}=R\otimes \Omega$, and $\cal G$ tends to zero sufficiently fast at $\omega = p^4 \rightarrow \pm \infty$. Also it may be easily found that for any $l$ the value of ${\cal M}_l$ is topological invariant if we deal with any compact momentum space.

\section*{Appendix C. Calculation of $\tilde{\cal N}_3$ for the $2+1$ D systems}

In this section we demonstrate how the value of the topological invariant $\tilde{\cal N}_3$  may be calculated.
Below we calculate $\tilde{\cal N}_3$ for the case, when the Green function has the form
 \begin{equation}
 {\cal G}^{-1}({\bf p}) = i\sigma^3\Big(\sum_{k}\sigma^{k} g_{k}({\bf p}) - i g_4({\bf p})\Big)\label{G12d}
 \end{equation}
 where $\sigma^k$ are Pauli matrices while $g_k({\bf p})$ and $g_4({\bf p})$ are the real - valued functions, $k = 1,2,3$. Let us define
 \begin{equation}
 {\cal H}({\bf p}) = \Big(\sum_{k}\sigma^{k} \hat{g}_{k}({\bf p}) - i \hat{g}_4({\bf p})\Big)
 \end{equation}
 where $\hat{g}_k = \frac{g_k}{g}$, and $g = \sqrt{\sum_{k=1,2,3,4}g_k^2}$. Then
 \begin{eqnarray}
\tilde{\cal N}_3 &=&  -\frac{1}{24 \pi^2} {\rm Tr}\, \int_{} \, {\cal H} d {\cal H}^+ \wedge d {\cal H} \wedge d {\cal H}^+\nonumber\\
&=&  \frac{1}{24 \pi^2} {\rm Tr}\, \int_{} \, {\cal H}^+ d {\cal H} \wedge d {\cal H}^+ \wedge d {\cal H} \nonumber\\
&=&  -\frac{1}{48 \pi^2} {\rm Tr}\,\gamma^5 \int_{} \, \tilde{\cal H} d \tilde{\cal H} \wedge d \tilde{\cal H} \wedge d \tilde{\cal H}\label{N3AH}
\end{eqnarray}
where
\begin{equation}
 \tilde{\cal H}({\bf p}) = \sum_{k=1,2,3,4}\gamma^{k} \hat{g}_{k}({\bf p}) = i{\rm diag}\,({\cal H},-{\cal H}^+)\gamma^4
 \end{equation}
 and $\gamma^k$ are Euclidean Dirac matrices in chiral representation, $\gamma^5$  in chiral representation is given by ${\rm diag}(1,1,-1,-1)$. This gives
  \begin{eqnarray}
\tilde{\cal N}_3 &=&  \frac{1}{12 \pi^2} \epsilon^{ijkl}\, \int_{} \, \hat{g}_i d \hat{g}_j \wedge d \hat{g}_k \wedge d \hat{g}_l
\end{eqnarray}
Let us introduce the parametrization
\begin{equation}
\hat{g}_4 = {\rm sin}\,\alpha, \quad \hat{g}_i = k_i\,{\rm cos}\,\alpha
\end{equation}
where $i=1,2,3$ while $\sum_{i}k^2=1$, and $\alpha \in [-\pi/2,\pi/2]$. Let us suppose, that $\hat{g}_4({\bf p})=0$ on the boundary of momentum space ${\bf p}\in \partial {\cal M}$. This gives
  \begin{eqnarray}
\tilde{\cal N}_3 &=&  \frac{1}{4 \pi^2} \epsilon^{ijk}\, \int_{\cal M} \, {\rm cos}^2 \alpha\, k_i\,  d\,\alpha  \wedge d k_j \wedge d k_k\nonumber\\ &=&  \frac{1}{4 \pi^2} \epsilon^{ijk}\, \int_{\cal M} \, k_i\,  d(\alpha/2+\frac{1}{4}{\rm sin}\,2\alpha)  \wedge d k_j \wedge d k_k \nonumber\\ &=& -\sum_l \frac{1}{4 \pi^2} \epsilon^{ijk}\, \int_{\partial{\Omega(y_l)}} \, k_i\,  (\alpha/2+\frac{1}{4}{\rm sin}\,2\alpha)   d k_j \wedge d k_k
\end{eqnarray}
In the last row $\Omega(y_l)$ is the small vicinity of point $y_l$ of momentum space, where vector $k_i$ is undefined. The absence of the singularities of $\hat{g}_k$ implies, that $\alpha \rightarrow \pm\pi/2$ at such points.

This gives
  \begin{eqnarray}
\tilde{\cal N}_3 &=&   -\frac{1}{2}\sum_l \, {\rm sign}(g_4(y_l)) \,{\rm Res}\,(y_l)
\end{eqnarray}
Following  \cite{Z2012} we use the notation:
\begin{eqnarray}
{\rm Res}\,(y) &=&  \frac{1}{8 \pi} \epsilon^{ijk}\, \int_{\partial \Omega(y)} \, \hat{g}_i  d \hat{g}_j \wedge d \hat{g}_k
\end{eqnarray}
It is worth mentioning, that this symbol obeys $\sum_l {\rm Res}\,(y_i)=0$.

Let us illustrate the above calculation by the consideration of the  particular example of the system with the Green function ${\cal G}^{-1} = i \omega - H({\bf p})$, where the Hamiltonian has the form
\begin{equation}
H = {\rm sin}\,p_1\, \sigma^2 - {\rm sin}\, p_2 \, \sigma^1 - (m^{(0)} + \sum_{i=1,2}(1-{\rm cos}\,p_i)) \, \sigma^3
\end{equation}
This gives
\begin{equation}
-i\sigma^3{\cal G}^{-1} = {\rm sin}\,p_1\, \sigma^1 + {\rm sin}\, p_2 \, \sigma^2 +  \omega \, \sigma^3 - i (m^{(0)} + \sum_{i=1,2}(1-{\rm cos}\,p_i))
\end{equation}
The boundary of momentum space corresponds to $\omega = \pm \infty$. We have
$$\hat{g}_4({\bf p}) = \frac{(m^{(0)} + \sum_{i=1,2}(1-{\rm cos}\,p_i))}{\sqrt{(m^{(0)} + \sum_{i=1,2}(1-{\rm cos}\,p_i))^2+ {\rm sin}^2\,p_1+ {\rm sin}^2\,p_2 + \omega^2}} $$
For example, for $m^{(0)}\in (-2,0)$ we have
\begin{eqnarray}
\hat{g}_4({\bf p}) & = & 0, \quad {\bf p}\in \partial{\cal M}\nonumber\\
\hat{g}_4({\bf p}) & = & -1, \quad \hat{g}_i({\bf p})  = 0\quad (k = 1,2,3),\quad  {\bf p} = (0,0,0)\nonumber\\
\hat{g}_4({\bf p}) & = & 1,\quad  \hat{g}_i({\bf p})  = 0\quad (k = 1,2,3),\quad  {\bf p} = (0,\pi,0) \nonumber\\
\hat{g}_4({\bf p}) & = & 1,\quad  \hat{g}_i({\bf p})  = 0\quad (k = 1,2,3),\quad  {\bf p} = (\pi,0,0)\nonumber\\
\hat{g}_4({\bf p}) & = & 1,\quad  \hat{g}_i({\bf p})  = 0\quad (k = 1,2,3),\quad  {\bf p} = (\pi,\pi,0)
\end{eqnarray}
Therefore, we get immediately
\begin{eqnarray}
\tilde{\cal N}_3 &=&  \frac{1}{2} - \frac{1}{2} (-1)-\frac{1}{2}(-1) - \frac{1}{2} = 1
\end{eqnarray}
In the similar way $\tilde{\cal N}_3  = -1 $ for $m^{(0)}\in (-4,-2)$ and $\tilde{\cal N}_3  = 0 $ for $m^{(0)}\in (-\infty,-4)\cup (0,\infty)$.

\end{document}